\documentclass{aa}  

\usepackage{graphicx}
\usepackage{lscape}
\usepackage{amssymb}
\usepackage{amsmath}

\usepackage{txfonts}

\usepackage{booktabs}
\usepackage{subfigure}
\usepackage{placeins}
\usepackage{verbatim}
\usepackage{adjustbox}
\usepackage[toc,page]{appendix} 
\usepackage{tabularx}
\usepackage{lscape}
\usepackage{textcomp}
\usepackage{gensymb}
\usepackage{multicol}
\usepackage[symbol]{footmisc}
\usepackage{tikz}
\usepackage{caption}
\usepackage{longtable}

\usepackage{natbib}
\usepackage{supertabular}

\usepackage{color}

\usepackage{perpage}
\MakePerPage{footnote}
\usepackage{lscape}

\usepackage{hyperref}  
\hypersetup{colorlinks=true,linkcolor=[rgb]{1.,0.2,0.2},citecolor=[rgb]{0.1,0.4,1.},filecolor=[rgb]{0.7,0.2,0.2},urlcolor=[rgb]{0.7,0.2,0.2}}

\newcommand{\bd}{\begin{displaymath}}
\newcommand{\ed}{\end{displaymath}}
\newcommand{\be}{\begin{equation}}
\newcommand{\ee}{\end{equation}}
\newcommand{\beaa}{\begin{eqnarray*}}
\newcommand{\eeaa}{\end{eqnarray*}}
\newcommand{\bea}{\begin{eqnarray}}
\newcommand{\eea}{\end{eqnarray}}

\def\macs1149{MACS 1149}

\begin{document} 

   \title{Unraveling the Lyman Continuum Emission of Ion3: Insights from HST multi-band imaging and X-Shooter spectroscopy}


\author{
U.~Me\v{s}tri\'{c}, \inst{\ref{unimi}, \ref{inafbo}}\
E.~Vanzella, \inst{\ref{inafbo}}\
A. Beckett,\inst{\ref{STScI}} 
M. Rafelski,\inst{\ref{STScI}, \ref{JHU}}
C. Grillo,\inst{\ref{unimi}, \ref{INAFmilano}}
M. Giavalisco, \inst{\ref{Amherst}}
M. Messa,\inst{\ref{inafbo}},
M. Castellano,\inst{\ref{INAF_Roma}}
F. Calura,\inst{\ref{inafbo}}
G. Cupani,\inst{\ref{INAF_Trieste}, \ref{IFPU}}
A. Zanella, \inst{\ref{INAF_Padova}}
P. Bergamini, \inst{\ref{unimi}}
M. Meneghetti, \inst{\ref{inafbo}}
A. Mercurio, \inst{\ref{unisa}, \ref{oacn}, \ref{infn-unisa}}
P. Rosati, \inst{\ref{uniFerrara}}
M. Nonino,
K. Caputi, \inst{\ref{Kapteyn}, \ref{DAWN}}
A. Comastri\inst{\ref{inafbo}}
}

\institute{
Dipartimento di Fisica, Universit\`a  degli Studi di Milano, via Celoria 16, I-20133 Milano, Italy \label{unimi}
\and
INAF -- OAS, Osservatorio di Astrofisica e Scienza dello Spazio di Bologna, via Gobetti 93/3, I-40129 Bologna, Italy \label{inafbo} 
\and
Space Telescope Science Institute, 3700 San Martin Drive, Baltimore, MD 21218, USA \label{STScI}
\and
Department of Physics and Astronomy, Johns Hopkins University, Baltimore, MD 21218, USA \label{JHU}
\and
INAF -- IASF Milano, via A. Corti 12, I-20133 Milano, Italy \label{INAFmilano}
\and
Istituto Nazionale di Astrofisica – Osservatorio Astronomico di Trieste, Via Tiepolo 11, Trieste, Italy \label{INAF_Trieste}
\and
IFPU – Institute for Fundamental Physics of the Universe, Via Beirut 2, 34014 Trieste, Italy \label{IFPU}
\and
Istituto Nazionale di Astrofisica, Vicolo dell'Osservatorio 5, 35137 Padova (Italy) \label{INAF_Padova}
\and
Department of Astronomy, University of Massachusetts Amherst, Amherst, MA 01002, USA \label{Amherst}
\and
INAF - Osservatorio Astronomico di Roma, via di Frascati 33, 00078 Monte Porzio Catone, Italy \label{INAF_Roma}
\and
 Università di Salerno, Dipartimento di Fisica "E.R. Caianiello",
Via Giovanni Paolo II 132, I-84084 Fisciano (SA), Italy \label{unisa} %
\and
INAF – Osservatorio Astronomico di Capodimonte, Via Moiariello 16, I-80131 Napoli, Italy \label{oacn} %
\and
INFN – Gruppo Collegato di Salerno - Sezione di Napoli, Dipartimento di Fisica "E.R. Caianiello", Università di Salerno, via Giovanni Paolo II, 132 - I-84084 Fisciano (SA), Italy. \label{infn-unisa} %
\and
Dipartimento di Fisica e Scienze della Terra, Università degli Studi di Ferrara, via Saragat 1, I-44122 Ferrara, Italy \label{uniFerrara}
\and
Kapteyn Astronomical Institute, University of Groningen, P.O. Box 800, 9700AV Groningen, The Netherlands \label{Kapteyn}
\and
Cosmic Dawn Center (DAWN), Copenhagen, Denmark \label{DAWN}
}

 \abstract{We provide a comprehensive analysis of Ion3, the most distant LyC leaker at $z=3.999$, using multi-band HST photometry (F390W, F814W, and F140W) and reevaluated X-Shooter spectroscopy.
 Deep HST F390W imaging enabled us to probe uncontaminated LyC flux blueward $\sim$880Å, while the non-ionizing UV 1500\AA/2800\AA~flux is probed with the F814W/F140W band. 
 High angular resolution allows us to properly mask low-$z$ interlopers and prevent contamination of measured LyC radiation.
We confirm the detection of LyC flux at SNR $\sim$3.5 and estimate the escape fraction of ionizing photons to be in the range $f_{\rm esc, rel}$ = 0.06 -- 1, depending on the adopted IGM attenuation.
Morphological analysis of Ion3 reveals a clumpy structure made of two main components, Ion3$_{\rm A}$ and Ion3$_{\rm B}$, with effective radii of 
R$_{\rm eff}$ $\sim$180 pc and 
R$_{\rm eff}$ < 100 pc, respectively, and a total estimated de-lensed area in the rest-frame 1600\AA~of 4.2~kpc$^{2}$. 
We confirm the presence of faint ultraviolet spectral features, including HeII$\lambda$1640, CIII]$\lambda$1907,1909 and [NeIII]$\lambda$3968, with rest-frame equivalent width EW(HeII) = (1.6$\pm$0.7)\AA\ and EW(CIII]) = (6.5$\pm$3)\AA.
From [OII]$\lambda$$\lambda$3726,3729 and  [CIII]$\lambda$1909/CIII]$\lambda$1906 doublets
we derive electron densities $n_{\rm e}^{\rm [OII]}$ = 2300$\pm$1900 cm$^{-3}$ and $n_{\rm e}^{\rm CIII]}$ > 10$^{4}$ cm$^{-3}$, corresponding to an ISM pressure log(P/k) > 7.90. Furthermore, we derive an intrinsic SFR(H$\alpha$) $\approx$ 77 M$_{\odot}$ yr$^{-1}$ (corresponding to $\Sigma_{\rm SFR} = 20$~M$_{\odot}$~yr$^{-1}$~kpc$^{-2}$ for the entire galaxy) and sub-solar metallicity $12+\rm log(O/H)$ = 8.02$\pm$0.20 using the EW(CIII]) as a diagnostic. The detection of [NeIII]$\lambda$3968 line and [OII]$\lambda$$\lambda$3726,3729, provide an estimate of the ratio [OIII]$\lambda$5007/[OII]$\lambda$$\lambda$3727,29 of O32 > 50 and high ionization parameter log$U$ > $-$1.5 using empirical and theoretical correlations. These measurements imply remarkably high ionization and density conditions produced by an ongoing bursty star formation, observed during a ionizing optically-thin phase of the ISM along the line of sight.
}

   
   \titlerunning{Ion3}
   \authorrunning{Meštrić et al.}
   \maketitle

%

\section{Introduction}
\label{sec:intro}

Directly detecting Lyman continuum radiation (LyC, $\lambda < 912Å$) emitted by star-forming galaxies at redshifts $3 < z < 4.5$ is a challenging task. After more than 20 years of searching for LyC-emitting galaxies using different approaches, such as spectroscopy and broad and narrow-band photometry, only a handful of LyC sources have been confirmed at these redshifts \citep[e.g.,][]{vanz_ion2, Shapley2016, Ion3_2018, steidel18, Fletcher2019, Rui2021, Rui2022, rivera19}. 
The current number of galaxies at these redshifts with confirmed LyC detection is not sufficient to build a robust and statistically significant sample required to characterize the properties of LyC-leaking galaxies and understand the mechanisms and conditions that control the LyC emissivity. Several factors limit our ability to successfully detect hydrogen-ionizing radiation emitted by star-forming galaxies.
First, LyC radiation is produced by O-type and other massive, short-lived stars and is easily absorbed by neutral hydrogen and dust in the ISM. This absorption makes it difficult for LyC radiation to be observable, and the situation is further exacerbated by the increasing intervening opacity of the circumgalactic (CGM) and intergalactic (IGM) mediums with increasing redshift \citep[][]{worseck14}.
Second, LyC flux can easily be contaminated by low-redshift sources, which further complicates ascertaining the nature of suspected LyC emission without high-angular resolution, multi-band imaging \citep[e.g.][]{Vanzella2010, Nestor2011}.
Overcoming these challenges and accurately measuring the escape fraction ($f_{\rm esc}$) of LyC radiation from galaxies is crucial for understanding how the Epoch of Reionization (EoR) progressed, as well as for the formation and evolution of galaxies.

In recent years, faint, low-mass galaxies have been suggested as the most likely main contributors to the global budget of ionizing photons \citep[e.g.,][]{Finkelstein2019, Simmonds2024}. The relative contribution of high-mass galaxies, however, still remains to be quantified,  and these sources still remain potential important candidates of substantial LyC emission \citep[e.g.,][]{Naidu2020_oligarch}. Besides, brighter galaxies are inherently easier to study, which is key if indirect diagnostics of LyC emissivity are used to investigate higher redshifts since, due to the high IGM opacity beyond $z \sim 4.5$ \citep{Inoue2008, Inoue2014}, it is impossible to directly probe the the ionizing UV radiation from galaxies at the EoR. For these reasons there are ongoing efforts to identify and study large samples of LyC emitters at low redshift that plausibly are good proxies, or ``analogs'' of $z > 6$ galaxies to inform on how to best trace LyC leakers at the EoR by means of robustly calibrated indirect diagnostics. Results from Lyman alpha (Ly$\alpha$) radiation transfer models suggest that a promising LyC tracer is a multiply-peaked Ly$\alpha$ emission line \citep[e.g.,][]{Verhamme2015, behrens14}.
Correlations between the LyC emissivity and the Ly$\alpha$ spectral properties have been investigated using samples of low-$z$ and high-$z$ LyC-leaking galaxies, and these studies have shown that a multi-peaked Ly$\alpha$ profile with a narrow emission peak close to the systemic redshift indicates optically-thin media to LyC radiation \citep[e.g.,][]{Verhamme2017, Izotov2018_LyC_0.46, vanz_sunburst, izotov21lowmass, rivera17}. Furthermore, \cite{Naidu2022} demonstrated that the properties of the Ly$\alpha$ line as velocity separation between Ly$\alpha$ peaks (V$_{\rm sep}$) and the central escape fraction of Ly$\alpha$ can discriminate between sources with high and low escape fractions of LyC photons.
Nevertheless, it is recognized that describing correlations between the LyC emissivity and other properties of the galaxies is a multi-parameter problem, and in order to indirectly identify likely LyC leakers and predict their escape fraction, different physical properties need to be considered simultaneously and described in terms of multi-varied distribution functions. For example, the results from a multivariate analysis by \citet{Jaskot2024_I} suggest that the EW of Lyman absorption features and the UV $\beta$ slope are the most important parameters, followed by escape fraction of Ly$\alpha$ photons ($f_{\rm esc, \rm Ly\alpha}$), dust excess (E(B--V)$_{\rm neb}$), star formation rate surface density ($\rm \Sigma_{SFR}$), and the O32 index. These conclusions are based on a sample of 35 local ($z \sim 0.3$) LyC leakers \citep[e.g., LzLCS,][]{Flury2022}.
A similar set of indirect tracers of LyC has also been studied for sources at high redshift ($z\sim3-4$), however larger samples of LyC leakers are required to better constrain the multivariate models that robustly predict $f_{\rm esc}$. Currently, the most effective diagnostics to  predict $f_{\rm esc}$ of $z \sim 3$ sources adopt dust attenuation, the UV $\beta$ slope, the O32 index, and $\Sigma_{\rm SFR}$ as the best physical parameters \citep{Jaskot2024_II}.

This work presents results from new HST multi-band imaging of Ion3 (PI: Meštrić, ID 17133) which, at redshift $z=3.999$, is the most distant confirmed LyC leaker to date \citep{Ion3_2018}. 
Ion3 shows strong LyC leakage and quadruple-peaked Ly$\alpha$ emission features, with a peak located at the resonance frequency, and a very blue ultraviolet continuum slope, $\beta$ slope of $-2.5\pm 0.25$. 
The photometric SED from VLT/HAWKI Ks and Spitzer/IRAC 3.6 µm and 4.5 µm bands (available from the Hubble frontier fields project, \citealt{Lotz_2017HFF} reveals a clear excess in the 3.6 µm channels ascribed to prominent H$\alpha$ line emission (with rest-frame EW $\sim$ 1000 Å).  From the same data,
a stellar mass of $\sim$1.5 $\times 10^{9} \rm M_\odot $, a SFR of $\simeq$140 $\rm M_\odot yr^{-1}$, and a young age of $\sim$10 Myr have been inferred \citep[][]{Ion3_2018}.

The depth and high-angular resolution, combined with the wavelength coverage of HST imaging allows us to measure $f_{\rm esc}$ and successfully check for and remove potential contamination by a low-$z$ interloper. Furthermore, we combine the HST imaging with archival spectroscopy (FORS2 and X-Shooter) and Spitzer/IRAC 3.6$\micro$m data to investigate the ISM conditions and mechanisms behind LyC leakage in Ion3 in depth.
This work is organized as follows: in Section \ref{sec:2}, we describe our HST data reduction procedure; in Section \ref{sec:3}, we detail the morphological, photometric, and spectroscopic analyses; in Section \ref{sec:4}, we define and evaluate the escape fraction of ionizing photons; Section \ref{sec:5}, provides results and discussion; and lastly, in Section \ref{sec:6}, we summarize and conclude our work.
Throughout this work, we adopt $\Lambda$-CDM cosmology $H_{0} = 68$ km s$^{-1}$ Mpc$^{-1}$, $\Omega_{\rm m} = 0.3$, and $\Omega_{\Lambda} = 0.7$.

\section{Data}
\label{sec:2}

\subsection{HST data}

The HST data used in this work consist of three bands of imaging using both the UV/Optical and NIR channels of the Wide Field Camera 3 (WFC3, \citealt{marinelli2024}), namely F390W, F814W and F140W. 
We use a long exposure time in F390W (25,540s across 9 exposures and 3 visits) to detect LyC radiation, from 890Å and blueward, from Ion 3. Shorter exposures are used to measure the non-ionizing UV flux at $\lambda \sim$1400Å to $\sim$1950Å) and stellar continuum morphology (8030s in F814W) and to improve the SED modeling to measure properties such as the stellar mass, as well as to check for low-$z$ interlopers (13,900s in F140W).
The full width half maximum (FWHM) of the point spread functions (PSFs) for the F390W and F8140W filters are very similar, $0.070''$ and $0.074''$ respectively. 
Therefore, PSF size matching was not performed between these two bands. Despite its expected FWHM of $\sim 0.141''$, which is notably different from other two observed bands, PSF matching to the F140W filter was not performed because this filter is only used for contamination checks in our analysis.

We applied a custom data reduction pipeline to this data, utilizing routines from the Drizzlepac software \citep{gonzaga12, hoffmann2021}, as well as LACOSMIC \citep{Dokkum2001} and custom routines similar to those used in several previous studies with WFC3 imaging \citep[e.g.][]{prichard2022, revalski2023, wang2024}. We first prepared the individual exposures using CALWF3, which produces the initial dark and flat-field corrections, debiasing, sink pixel rejection, and post-flash subtraction. Firstly, we flagged hot pixels using a variable threshold. The number of hot pixels varies as a function of location on the detector when using a constant threshold, due to imperfections in the corrections made to account for charge transfer efficiency (CTE) degradation. By fitting the number of hot pixels as a function of row number, we can then adjust the detection threshold such that the hot pixels are evenly distributed \citep{prichard2022}. Next, after masking out sources detected in each image, we estimated the background levels in the two chips, which were then equalized to remove any offsets between the amplifiers. 

We then proceeded to remove cosmic rays from the exposures. ASTRODRIZZLE was used for an initial run to detect cosmic rays. Read-out cosmic rays (ROCRs) were removed by flagging pixels within 5 pixels of these detected cosmic rays that are more than 2.75$\sigma$ below the background level. After this, we used LACOSMIC to remove the remaining cosmic rays from each frame. Finally, we used the UPDATEWCS routine to reset the astrometry of each exposure to that obtained from the Hubble guide stars, which are tied to the GAIA astrometry. 

Finally, we aligned the cleaned exposures onto the same WCS grid and combined them using routines from Drizzlepac. We used the TWEAKREG routine to match the positions of bright, compact sources across the image, and the ASTRODRIZZLE routine to combine exposures into our final science images. We first created `unaligned' drizzles by adding the exposures from each visit together. This results in one image in the F814W and F140W bands, and three F390W images. Using TWEAKREG, we matched compact sources in each exposure to their counterparts in the unaligned image and adjusted the wcs of the exposures to align with the drizzle. We then re-combined the aligned exposures to produce final visit-level drizzles. 

We attempted to match these visit-level drizzles to the GAIA catalog \citep{gaiacollaboration2023}, but there were too few unsaturated stars in our field for which we could accurately measure a centroid. We therefore instead used TWEAKREG to align our images to each other. For the F390W data, we measured the `tweak' required to align the three visit-level drizzles to each other, applied that tweak to the exposures making up each visit, and combined these aligned exposures into a final F390W image. Finally, we applied the same procedure to align the F814W and F140W images with the F390W: measuring the required shift to align the drizzles to the F390W image, applying it to the individual exposures, and then drizzling these into the final science images. 

This ensures excellent relative astrometry despite some uncertainty in the absolute astrometry. Each time an image is `tweaked' to align with another, the spread of the differences between the locations of the matched sources gives an indication of the uncertainty in the shift. We find that the F390W exposures are aligned to each other to an accuracy of $\approx$2 mas (0.05 native pixel), the F814W image is aligned to the F390W to within 10 mas (0.25 native pixel), and the F140W matches the F390W to within 13 mas (0.04 native pixel).

\subsection{VLT X-Shooter data}

Four hours X-Shooter observations Ion3 were executed during November 2017 with an average seeing conditions of 0.8 arcsec (prog. 098.A-0665, P.I. Vanzella), providing a final spectrum spanning the range 3400\AA\ $-$ 24000 \AA, with spectral resolution of $35 - 60$ km~s$^{-1}$. 
The data were partially presented by \cite{Ion3_2018}, focusing mainly on the Ly$\alpha$ line and possible detection of [OII]$\lambda$$\lambda$3726,3729 used to measure systemic redshift.  
In this work we carefully analyze the full spectral range reporting on additional emission 
ultraviolet features tracing the physical properties of the interstellar medium and the ionizing source. 
We refer the reader to \cite{Ion3_2018} 
(and references therein) for details about X-Shooter data reduction.

\section{Analysis}
\label{sec:3}
Our main scientific goals are threefold: (1) evaluate the escape fraction of ionizing photons, while eliminating possible low-$z$ contaminants, (2) investigate the morphology of the system and, (3) study the ISM properties of the LyC-emitting regions. 
The main results presented in this section are summarized in Table \ref{tab:1}.

\subsection{Lyman continuum emission and morphological analysis}
\label{sec:3.1}

\begin{figure*}
        \centering
        \includegraphics[width=\linewidth]{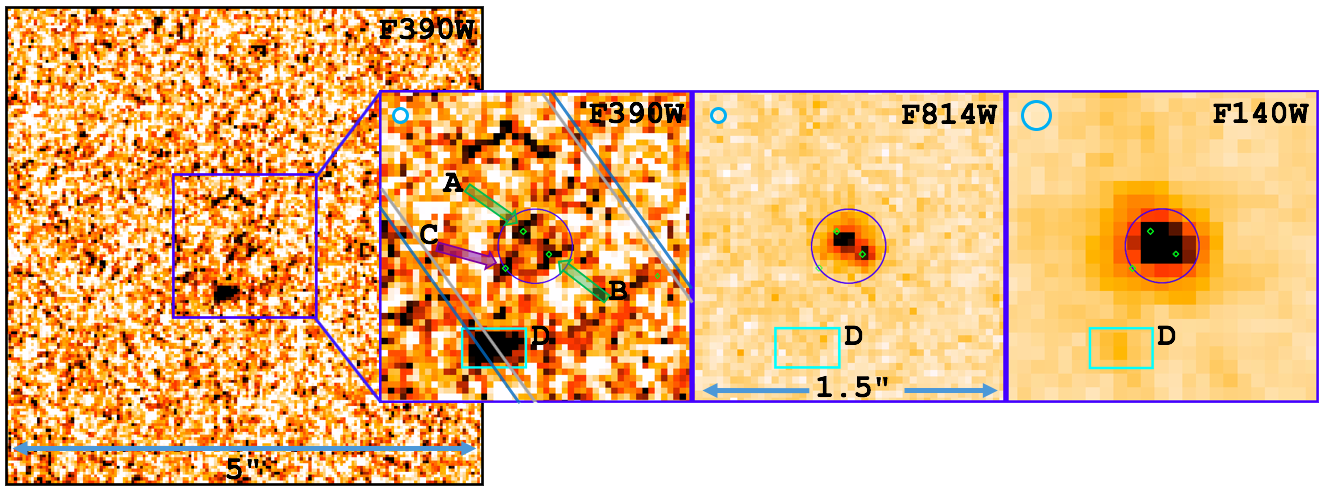}
        \caption{On the left side is the HST F390W band thumbnail image of Ion3, size $5''\times5''$. Three zoom-in cut outs of sizes $1.5''\times1.5''$ are shown on the right side and F390W, F814W and F140W bands with their expected PSFs in the upper left corner, blue circle. The HST F390W band probes LyC flux from 890Å and blueward (note that the F390W band misses the first $\sim20$Å of LyC flux). Green arrows point toward two clumpy structures (A and B) at $z=3.999$ where LyC flux is detected. The third clumpy structure, pointed to with a purple arrow (C), very close to the LyC leaking sources, is most likely a low-$z$ interloper, which is masked before the LyC flux is measured. Finally, the blue circle is adopted as the aperture to measure LyC flux, with a diameter of $0.36''$, while the cyan rectangle (D) encloses the source that probably contaminates the LyC flux detected from the FORS spectroscopy. The blue lines indicate the VLT FORS slit position, showing that source D enters the slit and likely contaminates the LyC flux detected in FORS spectroscopy. The grey lines show the VLT X-Shooter slit position. The second observed band is F814W and covers non-ionizing UV flux, where both LyC leaking clumps are detected. The third panel is the F140W band, where Ion3 is detected as a single bright blob, which is not resolved due to the lower spatial resolution of WFC3/IR. The three small green diamonds are centered on the clumpy sources A, B, and C in the F390W image, with their corresponding positions also shown in the F814W and F140W images. It is noticeable that clump A detected in F390W is slightly offset from clump A observed in F814W. }
        \label{hstIon3}
\end{figure*}

Figure~\ref{hstIon3} shows the LyC image at the redshift of Ion3 and the non-ionizing radiation at $\lambda \sim 1600$\AA\ and at $\lambda \sim 2800$\AA, from F390W, F814W and F140W bands, respectively. F814W data are much shallower than those in F390W (having $\rm 5\sigma$ magnitude limits mag$_{\rm lim(F814W)}\sim$25.41 and mag$_{\rm lim(F390W)}\sim$28.20, respectively).
We report the detection of a source, designated D in Figure 1, located 0.6 arcsec from Ion3.  Source D shows magnitudes of 28.00$\pm0.15$ and 27.50$\pm0.08$ in the F390W and F140W filters, respectively, but is not detected in F814W. 
We suspect that source D is a lower-redshift contaminant, previously 
unknown until this Hubble imaging analysis. Although we lack a redshift confirmation for source D, now it is clear that LyC signal detected from the ground-based FORS spectroscopic data presented in \citet{Ion3_2018} is probably contaminated by source D flux which leaks into the slit.
The deep F390W image probing LyC shows three tentative clumps at the location of Ion3 (A, B and C), two of which (A, and B) resemble the geometry of the double-blob morphology of Ion3 observed in the F814W image. Given the shallower F814W image, the non-detection of the object C does not necessarily rule out its LyC nature. However, we conservatively mask the source C in the following statistical analysis.

We perform photometric measurements using the {\tt Astropy} package and the photometry of astronomical sources with {\tt Photutils} \citep{Photutils2024}. 
The center of the aperture was defined by inspecting all three HST bands, and was placed between the detected clumps A and B; to check the robustness of our measurements, we moved the aperture center a two pixels around the selected center and concluded that the resulting photometry was unaffected.
Before measuring the LyC flux in F390W, we first mask the source C and construct a photometric curve of growth (CoG) with a series of apertures centered on the source. The CoG is computed by summing the flux in a predefined series of apertures to determine the optimal diameter for flux measurement (the peak at which the CoG reaches its maximum before entering the noise regime), which in our case is 12 pixels or $0.36''$ in diameter (blue circle, Figure \ref{hstIon3}); such aperture includes both A and B peaks in F390W and also most of the F814W flux. The mean background local sky level is measured within an annulus aperture ($r_{\rm in}$=7pix and $r_{\rm out}$=9pix).
The evaluated LyC flux and magnitude of Ion3 measured in the F390W band are $F_\nu$ = 1.07$\pm$0.3 $\times 10^{-31}$ erg/cm$^2$/s/Hz or $m_{F390W}$ = 28.8 $\pm$ 0.3 with SNR = 3.5.

\begin{figure}
        \centering
        \includegraphics[width=\linewidth]{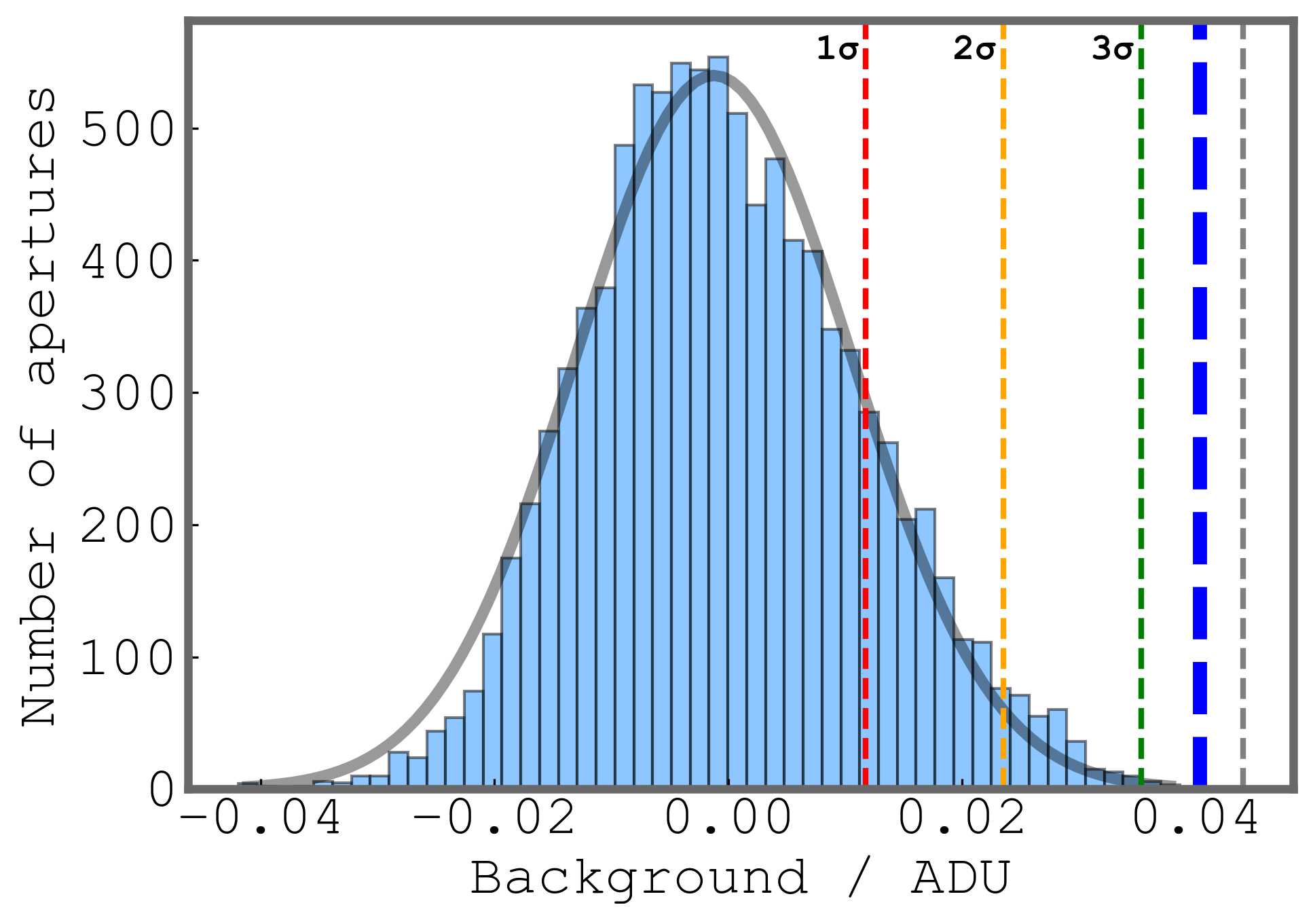}
        \caption{Histogram density plot of randomly placed 10,000 apertures over a $5''\times5''$ F390W image centered on Ion3. The black line is the Gaussian fit to the measured fluxes (in ADU) from 10k apertures, with a resulting mean value of 0.001 and a 1$\sigma$ of 0.0119, marked with a red dashed line (2$\sigma$ and 3$\sigma$ are marked with yellow and green dashed lines, respectively). The thick blue dashed line represents the background-subtracted, clean, non-contaminated LyC signal of Ion3, while thin gray dashed line is measured non-contaminated LyC signal before background subtraction. }
        \label{histo}
\end{figure}

To confirm the reliability and robustness of the measured LyC flux, we determine the background level in the parts of the images surrounding Ion3. As a first step, we create $5''\times5''$ thumbnails centered on Ion3 in all three observed HST bands. Since our aim is to probe the background around Ion3 in the F390W band, we detect all sources in F814W and F140W, adopting a detection threshold of 5$\sigma$. In the second step, all positions of detected sources in F814W and F140W are masked (in total 3 sources) in the F390W image to reduce contamination. Finally, we probe the background by randomly placing 10,000 overlapping apertures, of the same size used to measure the reported flux ($0.36''$ in diameter), over the $5''\times5''$ F390W thumbnail image.
The resulting background flux measurements are presented in the histogram density plot (Figure \ref{histo}); we confirm that the emission observed at the location of Ion3 is significant, at 3.5$\sigma$ level. 
Finally, the magnitude errors are evaluated by adopting the standard expression \texttt{mag\_error} = 1.086 $\times$ \texttt{err\_flux} /\texttt{flux}, where the term 
\texttt{err\_flux} is the standard deviation derived from 10,000 background apertures, and \texttt{flux} is the measured flux from Ion3 within the defined aperture.  
The same approach was applied for magnitude measurement in F814W band whose resulting magnitude is $m_{F814W}$ = 24.20 $\pm$ 0.07.

\begin{figure}
        \centering
        \includegraphics[width=\linewidth]{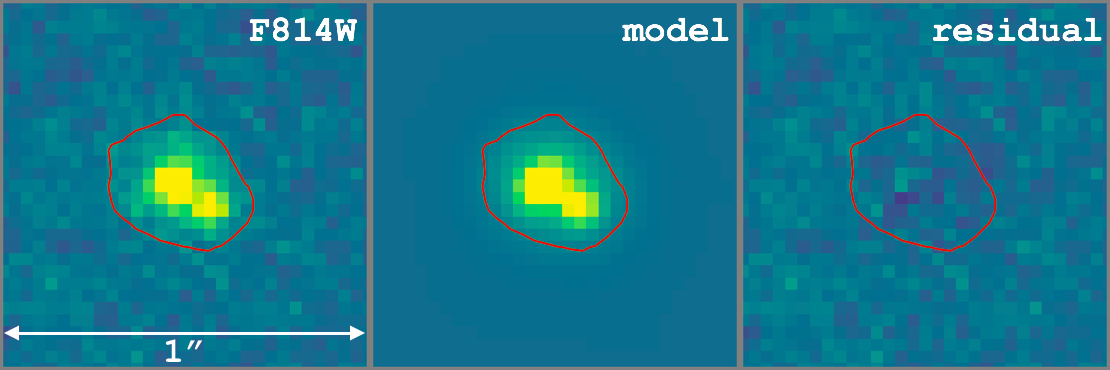}
        \caption{Results from {\tt Galfit} modeling. The first panel shows the original F814W image of Ion3 covering the non-ionizing UV flux with 2$\sigma$ contour in red, also overlaid in other two panels. The second panel is the resulting model from {\tt Galfit}, and the third panel is a residual map.}
        \label{gal}
\end{figure}

To measure the effective radius of both clumps, we applied the method described in \cite{M2022_clumps}. The two-dimensional modeling of the clump light profiles was performed using {\tt GALFIT} \citep{galfit_1, galfit_2}, resulting in estimated physical sizes for both components (Ion3$_{\rm A}$ and Ion3$_{\rm B}$). The sizes were derived from HST F814W imaging (probing UV rest-frame flux) after fixing the Sérsic index to 0.5 and leaving magnitude and half-light radii as free parameters. The resulting effective radii are 
R$_{\rm eff}$ $\sim$ 180$\pm90$ pc for Ion3$_{\rm A}$ and an upper limit of R$_{\rm eff}$ <100 pc for Ion3$_{\rm B}$.
The estimated sizes for both clumps are corrected for total magnification $\mu_{\rm tot}=1.3$ \citep{Caminha2016}. The {\tt GALFIT} model and residuals are shown in Figure \ref{gal}.

\subsection{X-Shooter spectroscopy}
\label{sec:3.2}

Our X-Shooter data includes a total of 4 hours of exposure time, divided into 4 observation blocks (OBs); the spectral range covered is  3000--5595Å (600--1119Å rest frame) by the UVB arm, 5595--10240Å (1119--2048Å rest frame) by the VIS arm, and 10240--24800Å (2048--4960Å rest frame) by the NIR arm. We will focus our analysis mostly on the spectra from the VIS and NIR arms, since the wavelength coverage of the UVB arm has been observed with deeper FORS spectroscopy and presented in \cite{Ion3_2018}. 
We combine the 4 OBs by creating an average weighted 2D stacked spectrum using the $\tt IRAF$ $\tt IMCOMBINE$ task (both for the VIS and NIR data).

\subsubsection{X-Shooter VIS-arm}
\label{sec:3.2.1}

\begin{figure*} 
        \centering
        \includegraphics[width=\linewidth]{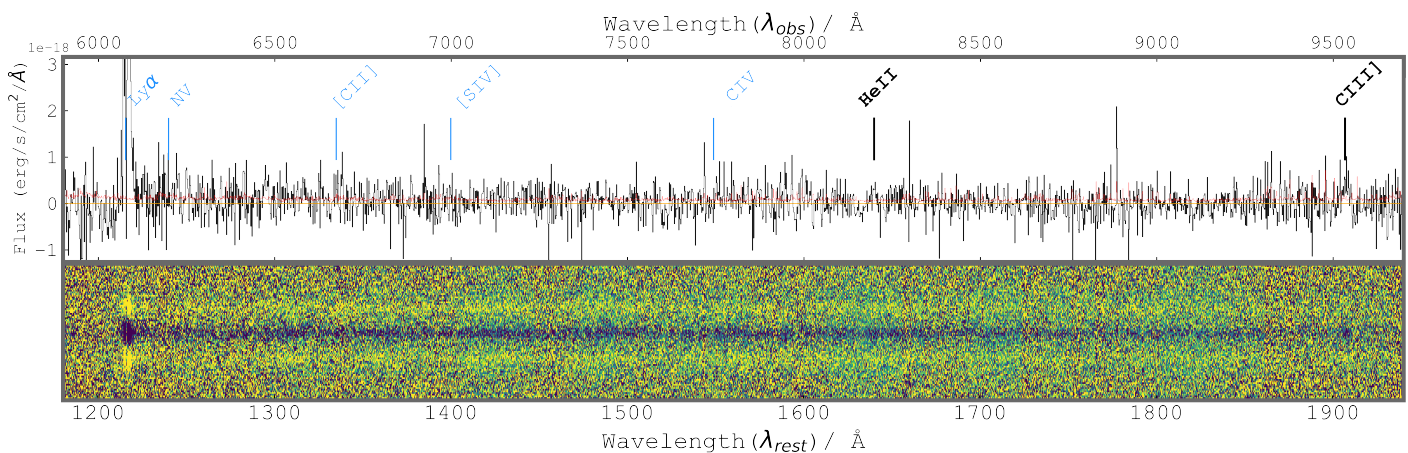}
        \caption{The X-Shooter VIS-arm 1D spectra (upper panel) and 2D spectra (bottom panel) covering the rest-UV wavelengths from $\sim$ 1180Å -- 1940Å. The spectrum (black) is binned by a factor of 9 to increase the signal-to-noise of the continuum and other spectroscopic features, while 1$\sigma$ uncertainty is presented in red color and orange line shows 0 flux level.  The displayed 1D spectrum is continuum subtracted. The location of the reported and analyzed emission features (HeII$\lambda$1640 and CIII]$\lambda$$\lambda$1907,1909) are marked with thick black markers, while the positions of the other lines reported in \cite{Ion3_2018} from FORS spectroscopy (Ly$\alpha$, NV$\lambda$1240, [CII]$\lambda$1335.71, [SIV]$\lambda$$\lambda$1393,1402, CIV$\lambda$$\lambda$1548,1550) are marked with thin blue markers. 
}
        \label{VIS_spec}
\end{figure*}

\begin{figure} 
        \centering
        \includegraphics[width=\linewidth]{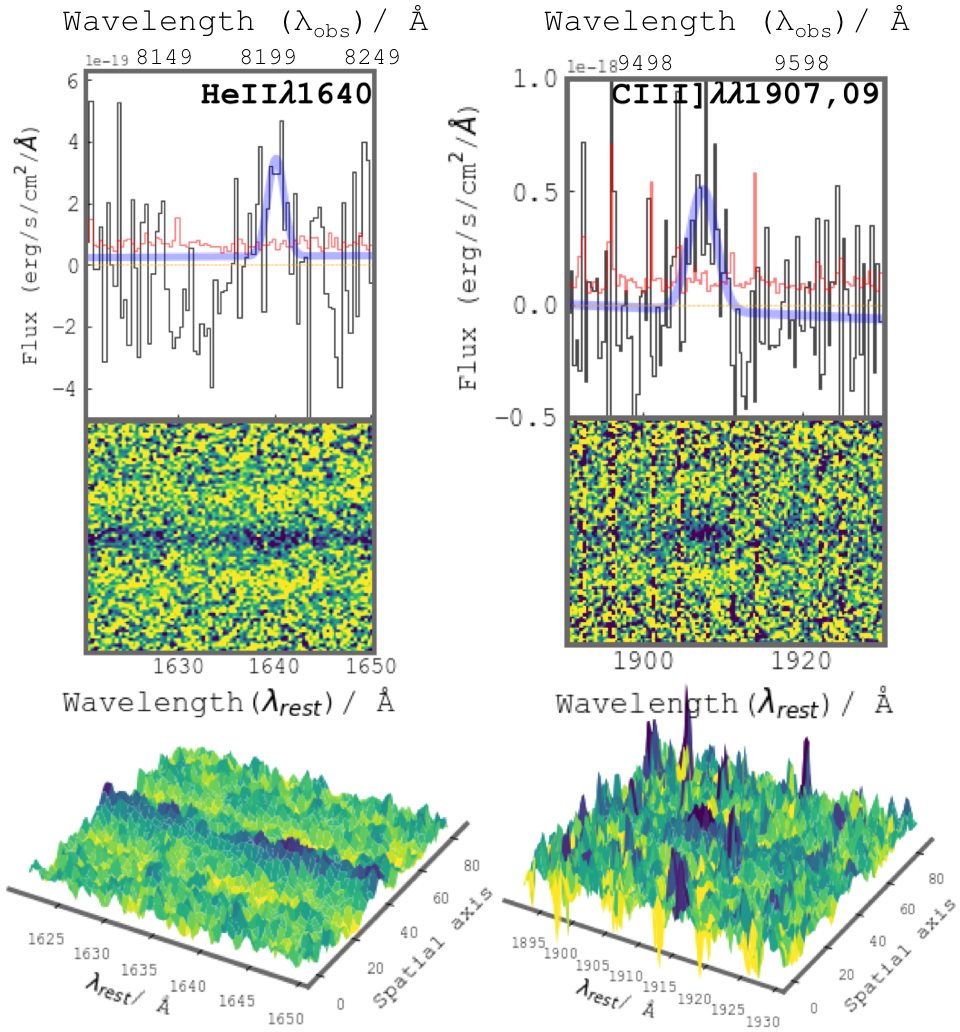}
        \caption{Zoom-ins show two emission features detected in the VIS arm of the X-Shooter spectra.
The left and right panels consist of the 1D continuum subtracted (upper), 2D (middle), and 3D (bottom) spectrograms of HeII$\lambda$1640 with characteristic P-Cygni line profile and semi-forbidden non-resolved CIII]$\lambda$$\lambda$1907,1909 line.
The fitted Gaussian emission line is overplotted in blue. The 1D spectra in both panels is shown in black, error spectrum in red and zero level flux with orange line. 
Sharp peaks notable in 1D around the CIII]$\lambda$$\lambda$1907,1909 line belonging to the sky residuals are also visible in 2D and 3D spectrograms. The 1D and 2D spectrograms are displayed in rest and observed wavelength frame.}
        \label{vis_lines}
\end{figure}

In order to increase the SNR of the VIS spectrum, we binned the 2D spectrum (whose initial resolution element is 0.2 Å/pix) by a factor of 9 in the dispersion direction using the $\tt IRAF$ $\tt BLKAVG$ task. This produced a binned 2D spectrum with a 1.8 Å/pix resolution element. Subsequently, the one-dimensional spectrum was extracted within a window of 1.6 arcsec centered on the spatial position of Ion3 in the slit, traced by its continuum. The result is shown in Figure \ref{VIS_spec}. After a careful inspection of the 1D and 2D X-Shooter VIS binned spectra, in addition to Ly$\alpha$, NV$\lambda$$\lambda$1238,1242, [CII]$\lambda$1335.71, CIV$\lambda$1550 lines \citep{Ion3_2018}, we detected HeII$\lambda$1640 and CIII]$\lambda$$\lambda$1907,1909 features.
Before measuring the flux and EW of newly detected emission lines, we fit the continuum and subtract the resulting fit from the original spectrum, creating a continuum-subtracted spectrum. The continuum fitting is done using the $\tt IRAF$ task $\tt CONTINUUM$, where we used a second-order polynomial, avoiding parts of the spectrum where prominent emission and absorption features are expected. The flux and equivalent width (EW) of detected emission lines are evaluated by fitting a Gaussian function to the observed spectral features.  This fitting procedure utilizes the $\tt LevMarLSQFitter$ class from the $\tt astropy$ library (unless otherwise noted). Uncertainties associated with the measured flux and EW are propagated from the covariance matrix output by the fitting algorithm. This matrix provides information on both the uncertainties and the correlations between the fitted parameters.
The resulting fluxes and EWs are $F_{\rm HeII}$=3.4$\pm2.5$$\times10^{-18}\rm erg/s/cm^2$, EW(HeII) = 1.6$\pm0.7 Å$ and $F_{\rm CIII]}$=1.12$\pm0.49$$\times10^{-17}\rm erg/s/cm^2$, EW(CIII]) = 6.5$\pm3 Å$. 
The fitted emission features and their corresponding 1D, 2D, and 3D spectrograms are shown in Figure \ref{vis_lines}.

\subsubsection{X-Shooter NIR-arm}
\label{sec:3.2.2}

\begin{figure*}
        \centering
        \includegraphics[width=\linewidth]{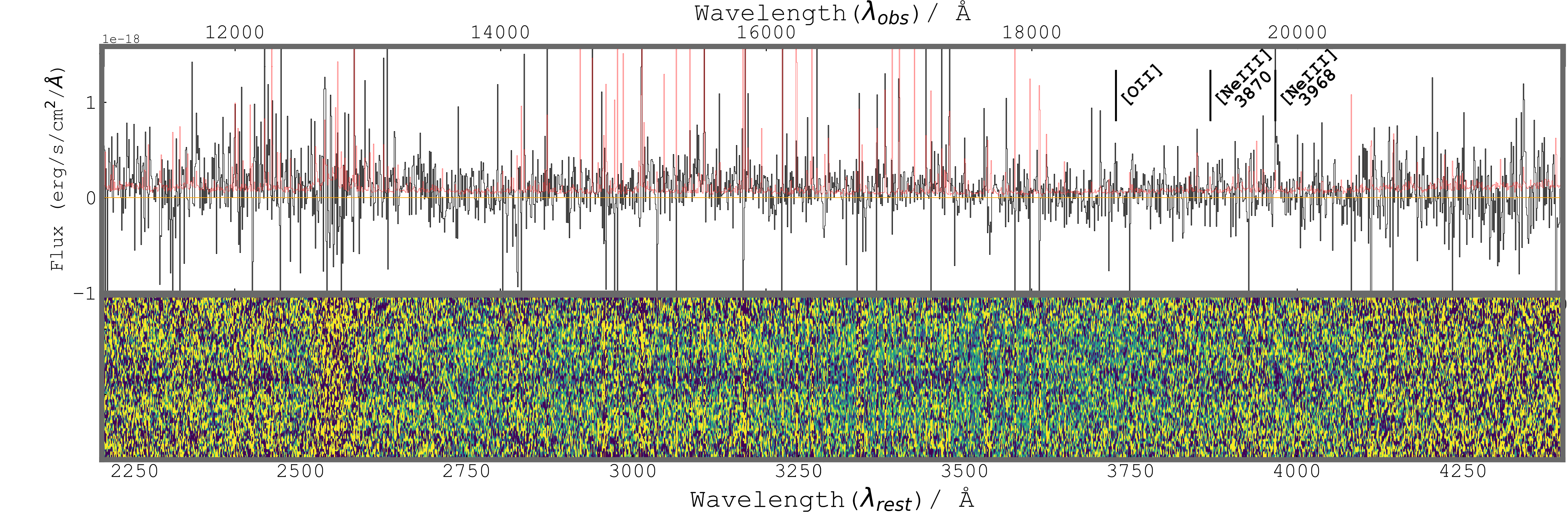}
        \caption{Same as Figure \ref{VIS_spec}, but for the X-Shooter NIR-arm, showing [OII]$\lambda$$\lambda$3727,3729 doublet and [NeIII]$\lambda$3968 line.}
        \label{NIR_spec}
\end{figure*}

\begin{figure} 
        \centering
        \includegraphics[width=\linewidth]{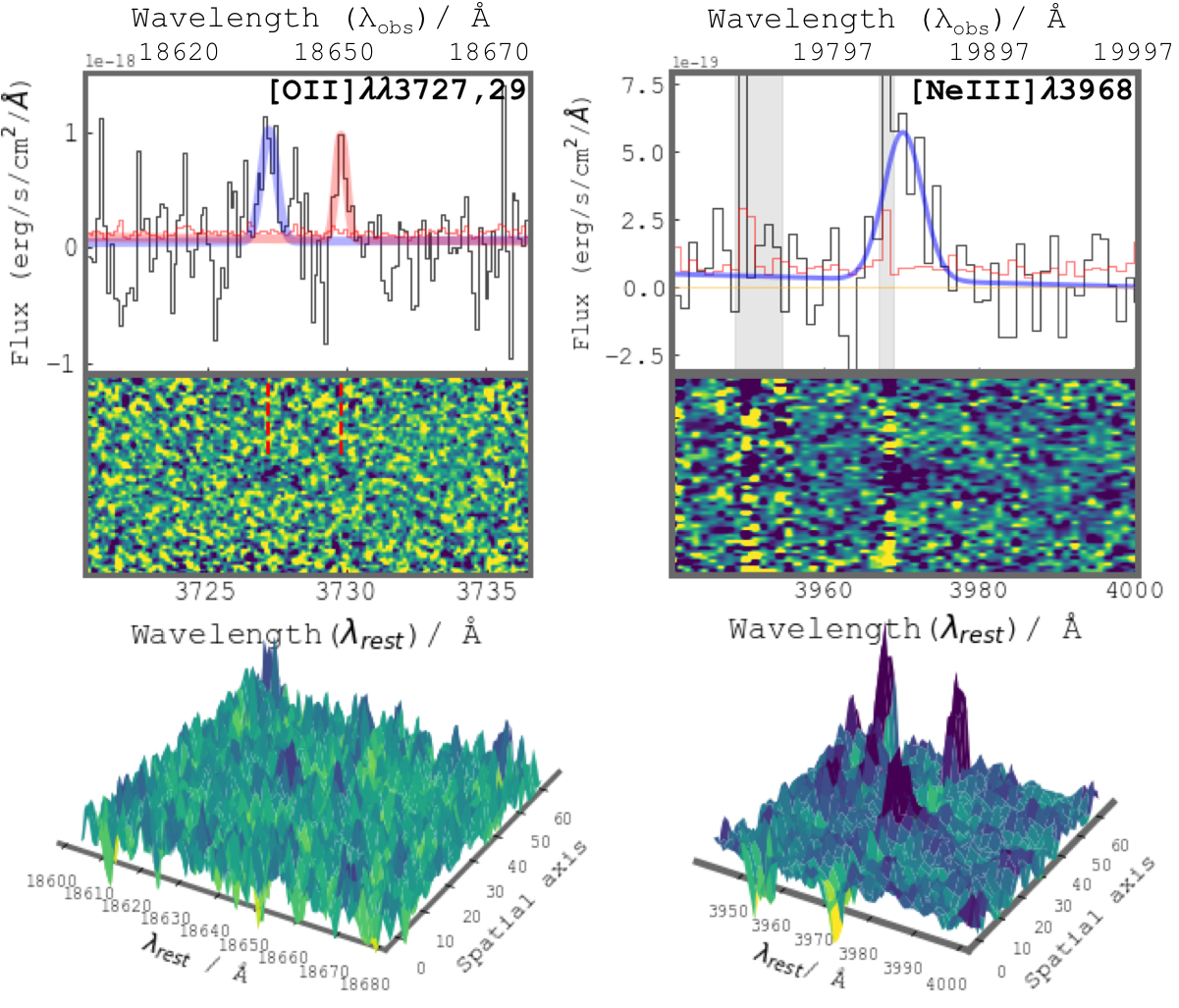}
        \caption{Same as Figure \ref{vis_lines}, but for the emission [OII]$\lambda$$\lambda$3727,3729 doublet and [NeIII]$\lambda$3968 line detected in X-Shooter NIR spectrum. For the [OII]$\lambda$$\lambda$3727,3729 doublet, the fitted Gaussian emission lines are overplotted in blue and red, respectively. The dashed red markers in [OII] 2D spectrogram point to the faint doublet at wavelengths [OII]$\lambda$$\lambda$3727,3729. Spectrum in both panels are presented in black while error in red color and zero level flux with orange line. Note that the [NeIII]$\lambda$3968 line is partially blended with the skyline (gray shaded region of $\sim$6Å width). Skyline is not taken into account when fitting a Gaussian curve and calculating the integrated line flux. The 1D and 2D spectrograms are displayed in the rest and observed frame wavelengths.}
        \label{nir_lines}
\end{figure}

NIR arm data is analyzed using a similar methodology as the VIS data described in the previous subsection. Upon preliminary visual examination, the final stacked spectra observed from the NIR arm, covering the $\sim$2050Å--4960Å rest-frame range, shows a very faint trace  of  continuum, visible down to $\sim$3500Å. 
Faint [OII]$\lambda$$\lambda$3727,3729 doublet with $F_{3727}$=3.37$\pm1.66$ $\times10^{-18}\rm erg/s/cm^2$ and $F_{3729}$=2.07$\pm1.38$ $\times10^{-18}\rm erg/s/cm^2$ flux is detected and confirms the systemic redshift of $z_{spec}=3.999 \pm 0.001$
reported by \citet{Ion3_2018}.
We calculated the integrated flux from the lines by fitting the Gaussian line profile (using the same method described in Section \ref{sec:3.2.1}), excluding the continuum-fitting procedure, being the continuum extremely faint and negligibly contributing in this case.
Furthermore, we bin the NIR data by a factor of 9, translating the spectrum from 0.5 Å/pix to 5.4 Å/pix resolution element to check for other fainter features, and show the resulting spectrum in Figure \ref{NIR_spec}. 
The [NeIII]$\lambda$3968Å feature is clearly detected and partially blended with a skyline. To quantify the integrated flux of this line, we followed a methodology similar to that adopted for the [OII] doublet and other VIS lines. We first mask the skyline region (6\AA\ wide region) to prevent its interference with the Gaussian fitting process and after we measure the integrated flux, as depicted in Figure \ref{nir_lines}.
The resulting flux is $F_{\rm [NeIII]}$=$(1.56 \pm0.54)\times10^{-17}\rm\ erg/s/cm^2$.

\section{Fraction of Escaped LyC photons}
\label{sec:4}

We derive the relative escape fraction of Lyman continuum photons ($f_{\rm esc,rel}$) following \cite{Vanzella2012, Ion3_2018}:

\begin{equation}
\label{eq:1}
f_{\rm esc, rel}=\frac{L_{\nu}(1500)/L_{\nu}(LyC)}{F_{\nu}(1500)/F_{\nu}(LyC)}\exp{}(\tau_{IGM}^{LyC}),
\end{equation}

where $L_{\nu}(1500)/L_{\nu}(LyC)$ represents the intrinsic flux density ratio of non-ionizing UV radiation (in our case from $\lambda_{\rm rest}$ = 1400 -- 1940 Å) and LyC photons, while ${F_{\nu}(1500)/F_{\nu}(LyC)}$ refers to the observed flux density ratio. The $\tau_{IGM}^{LyC}$ takes into account the attenuation of LyC photons caused by the neutral Hydrogen atoms along the line of sight.

From the HST F390W and F814W imaging the LyC flux blueward from 880Å (the F390W band probes rest-frame range $700 - 880$\AA, with a pivotal wavelength of 800\AA) and non-ionizing flux at 1500Å is measured 
F${_\nu}^{LyC}$ =1.1$\pm0.3$ $\times10^{-31}\rm erg/cm^2/s/Hz$ and F${_\nu}^{1500}$ = 7.6$\pm0.5$ $\times 10^{-30}$erg/cm$^2$/s/Hz, which correspond to flux density ratio ${F_{\nu}(1500)/F_{\nu}(LyC)}$ = 69$\pm19$.
On the other side, the intrinsic luminosity ratio is sensitive to the choice of stellar population synthesis models and assumptions about key galaxy properties, including metallicity, age, star formation history, and initial mass function. This dependence can cause the $L_{\nu}(1500)/L_{\nu}(LyC)$ ratio to vary significantly, from approximately 1.5 to 9, resulting in substantial uncertainty in the calculated escape fraction. 
Therefore, in order to put more stringent constraints on $L_{\nu}(1500)/L_{\nu}(LyC)$ we rely on the expression from \cite{chisholm19} that describes the relationship between stellar population age, metallicity, and the ratio between intrinsic ionizing and non-ionizing flux density:

\begin{equation}
\label{eq:X}
\frac {L(800)}{L(1500)}=(3.5\pm0.3) \times 10^{-({0.14\pm0.01}) \left (\frac{\rm Age}{\rm 1~Myr}\right)-(0.21\pm0.02) \left (\frac{\rm Z_{*}}{\rm 1Z_{\odot}}\right)}, 
\end{equation}

The presence of the NV$\lambda$1240 feature that shows a P-Cygni profile with clear absorption and emission \citep[FORS spectra,][]{Ion3_2018} is a strong indicator of young stellar populations ($<$5Myr) \citep[][]{chisholm19}.
The strength of the peak/through ratio can serve as an age indicator, where a higher ratio goes with younger ages.
For Ion3, the estimated NV$\lambda$1240 peak/through ratio is $\sim$2.3, which corresponds to an age of $\sim5$Myr, and with a given metallicity of 12+log(O/H)=8.02$\pm0.20$ or $\sim$0.2Z$_{\odot}$, (see Section \ref{sec:5.4}) we derive an intrinsic luminosity ratio of $L(1500)/L(LyC)$ = 1.6$\pm0.9$.

In order to fully sample the IGM transmission expected in F390W at $z$ = 4.0, we simulate 10,000 mock sightlines using the prescription from \citet{bassett2021}. To summarize, the column density of HI in the IGM is modeled by randomly adding HI absorbers with column densities between 10$^{12}$ and 10$^{21}$ cm$^{-2}$ to the mock sightlines using a column density distribution function given by \citet{steidel18}. At wavelengths redder than 912\AA, these produce Voigt profiles with doppler widths randomly sampled from \citet{hui1999}. The first 32 Lyman series transitions are modeled. At bluer wavelengths, the absorption is proportional to N$_{\rm HI}\lambda^{3}$ \citep[e.g.][]{Osterbrock1974}. 
In our simulations we do not include the CGM attenuation since the Ly-alpha multi-peak line suggests a low-column-density of CGM (see Section \ref{sec:4.1}).
\begin{figure} 
        \centering
        \includegraphics[width=\linewidth]{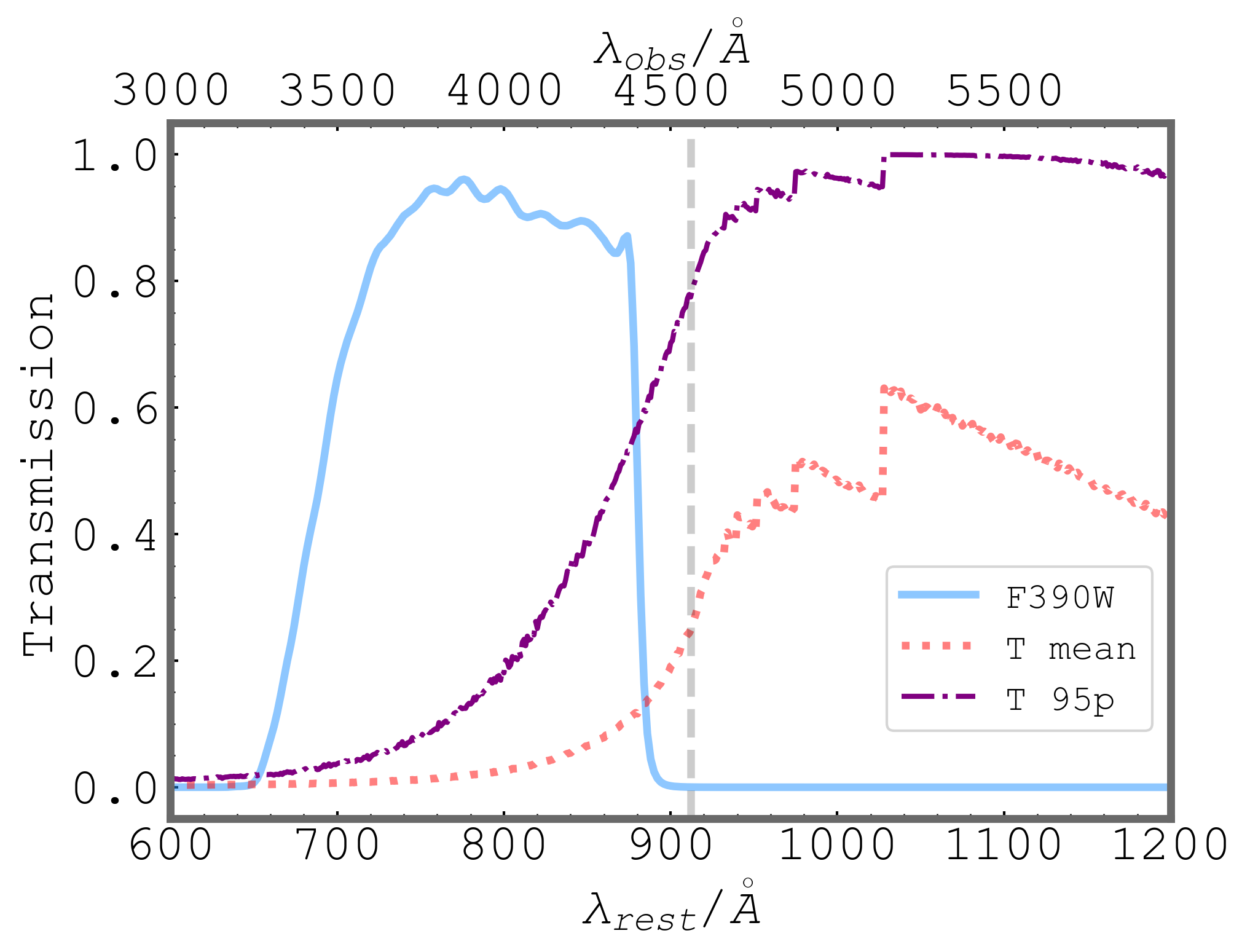}
        \caption{The mean IGM transmission as a function of wavelength over 10,000 lines of sight is shown as a red dotted line. The purple dash dotted line is 95 percentile IGM transmission and the blue solid line represents the F390W filter curve. A vertical gray dashed line marks the beginning of LyC region at 912Å.}
        \label{trans}
\end{figure}
The resulting IGM transmission is shown as a function of wavelength in Figure \ref{trans}. We then integrate this through the F390W filter to model the probability distribution for the fraction of LyC photons from Ion3 that will pass through the IGM. This procedure is repeated for all 10,000 sightlines, with the resulting $\tau_{IGM}^{LyC}$ 
distribution shown in Figure~\ref{Tigm_dist}.

To evaluate the relative escape fraction of ionizing photons, we adopt the previously estimated $L(1500)/L(LyC)$ = 1.6$\pm0.9$, ${F_{\nu}(1500)/F_{\nu}(LyC)}$ = 69$\pm19$ and the range of values for $\tau_{IGM}^{LyC}$ shown as the blue part of histogram plot in Figure \ref{Tigm_dist}.
The values of $\tau_{IGM}^{LyC}$ which produce $f_{\rm esc, rel} > 1$ (red histogram marked in Figure \ref{Tigm_dist}) are not considered in the analysis.
The corresponding distribution of $f_{\rm esc}$ is 
shown in the histogram plot of Figure~\ref{fesc_dist}, and spans the range $f_{\rm esc, rel}$ = 0.06 -- 1, depending on the $\tau_{IGM}^{LyC}$.
Finally it is worth noting that, given the already large distribution of $f_{\rm esc}$ values, we did not include LyC flux measurement uncertainty in the $f_{\rm esc, rel}$ distribution shown in Figure 10.

\begin{figure} 
        \centering
        \includegraphics[width=\linewidth]{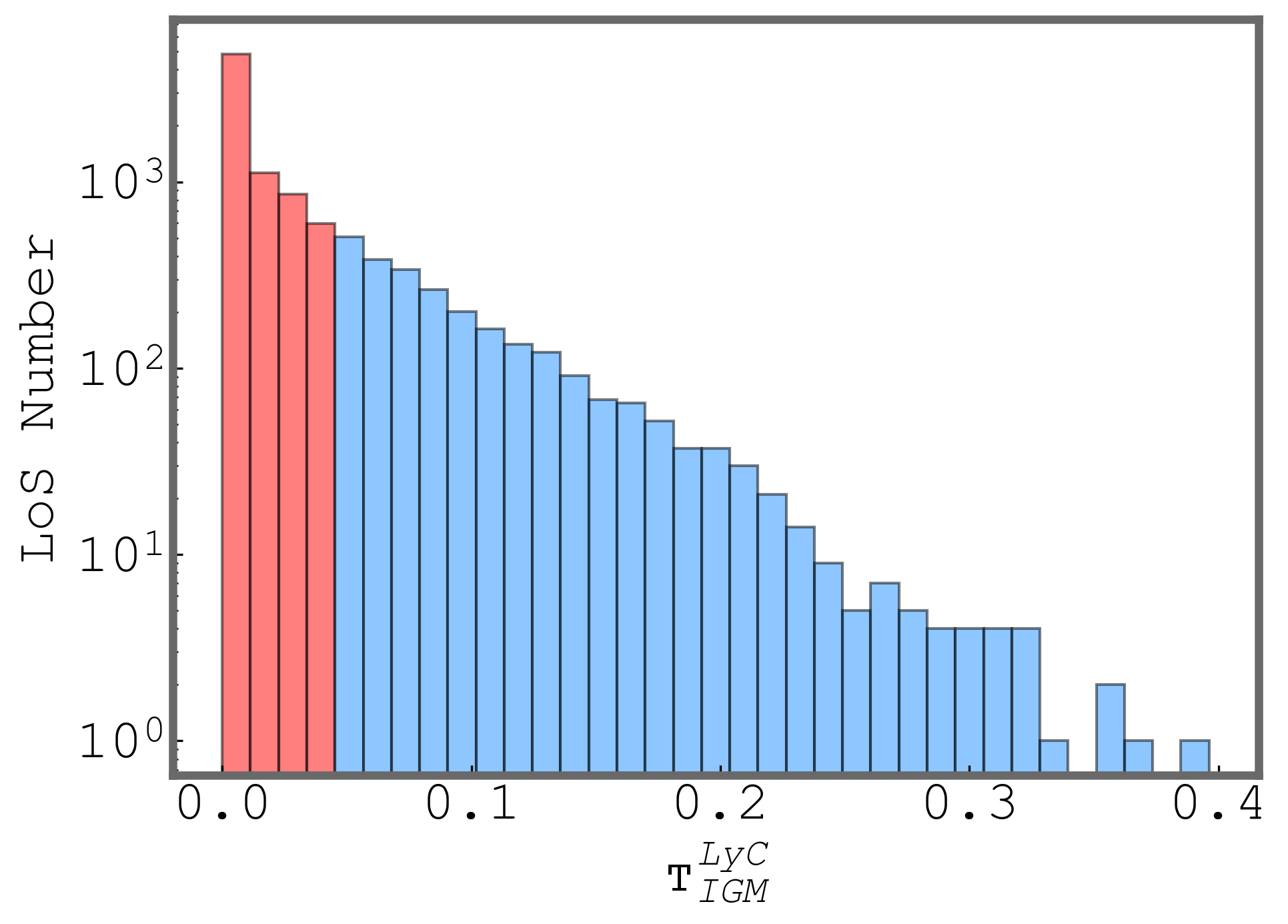}
        \caption{Resulting distribution of IGM transmission for 10,000 sightlines. The blue part of the histogram marks the IGM transmission distribution of those sightlines that in combination with adopted intrinsic and observed flux ratio result in $f_{\rm esc,rel}$=0.06--1. The IGM transmissions shown in red result in $f_{\rm esc,rel}$>1.}
        \label{Tigm_dist}
\end{figure}

\begin{figure} 
        \centering
        \includegraphics[width=\linewidth]{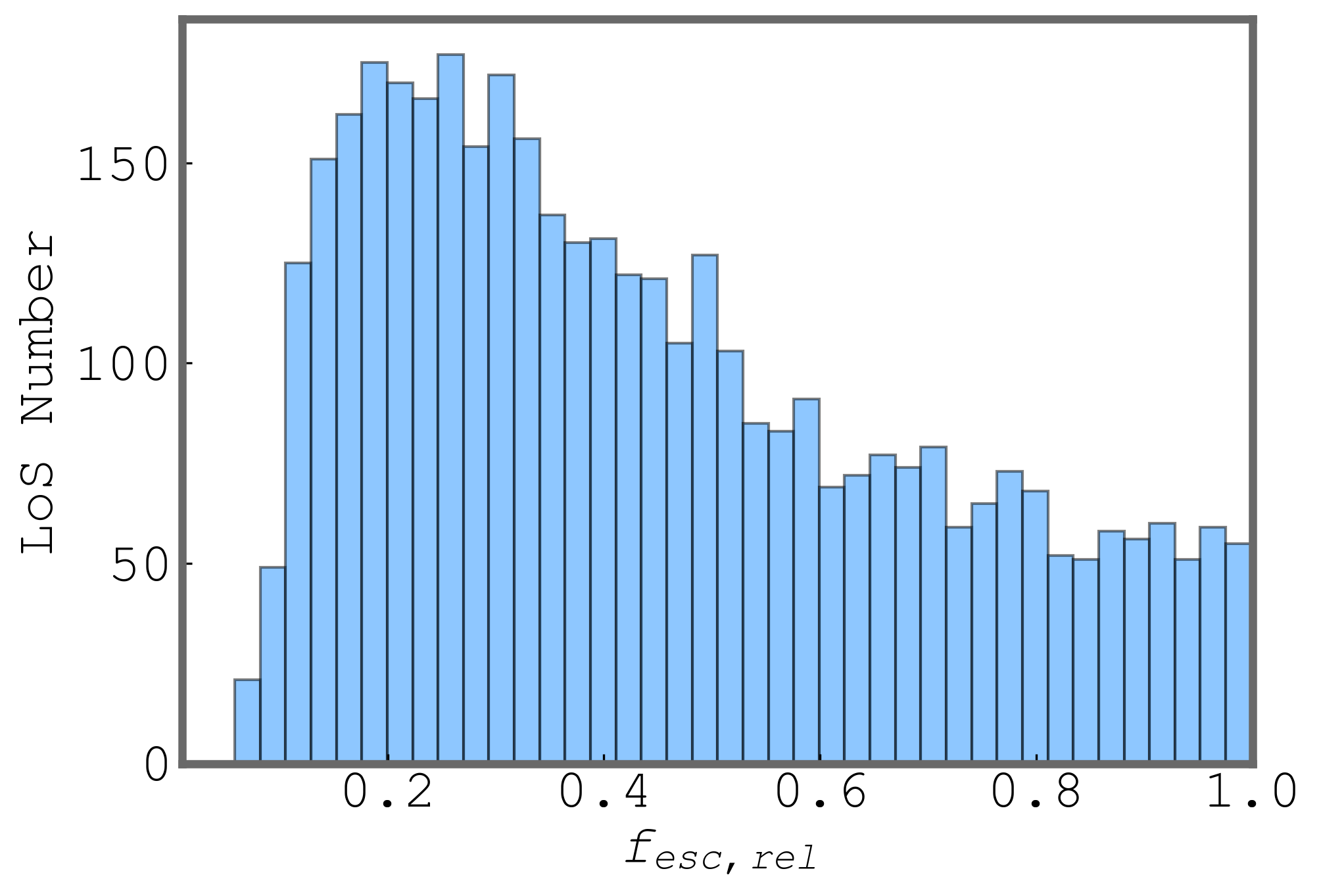}
        \caption{The distribution of evaluated $f_{\rm esc,rel}$ for Ion3 LyC leaker. The presented distribution only involve the $\tau_{IGM}^{LyC}$ sightlines (the blue part of the histogram in Figure \ref{trans}) for which $f_{\rm esc,rel}$=0.06--1. }
        \label{fesc_dist}
\end{figure}

\section{Results and Discussion}
\label{sec:5}

\subsection{The electron density (n$\rm _e$) and ISM pressure (log(P/k))} 
\label{sec:5.1}

To constrain the underlying physical conditions within the ISM (electron density, electron temperature, pressure, metallicity, and ionization parameter), a variety of robust line diagnostics have been proposed. One of the most commonly used density diagnostics is the forbidden fine-structure [OII]$\lambda$$\lambda$3727,3729 optical line doublet.
The [OII] doublet is produced by collisional excitations and de-excitations, therefore the [OII]$\lambda$$\lambda$3727/3729 line ratio is sensitive to the electron density of the environments with T$\sim$1-2$\times10^{4}$ K, recognized as star-forming regions \citep{Osterbrock.book.}.
Furthermore, the [OII]$\lambda$$\lambda$3727/3729 line ratio is a useful electron density diagnostic over the $40 - 10000$ cm$^{-3}$ range, with errors up to 0.4dex due to electron temperature dependency \citep[for more details, the reader is referred to][]{Kewley2019}.

We adopted the method of calculating n$_{\rm e}$ described in \cite{Sanders2016}, where electron density is expressed as a function of the line ratio:

\begin{equation}
\label{eq:3}
n_{\rm e}(R)= \frac{cR - ab}{a - R},
\end{equation}

where [OII]$\lambda$$\lambda$3727/3729 line ratio is denoted as $R$ and $a=0.3771$, $b=2,468$ and $c = 638.4$ are the best fit coefficients.
It is important to highlight that $n_{\rm e}$ measurements are temperature dependent which can introduce errors of about $\sim20\%$ \citep{Sanders2016}.  In our case  we adopt T$_{e}$=10$^4$K \citep{Osterbrock1974}.
The uncertainty due to the relatively low signal to noise ratio of the [OII] components dominates the electron density measurement, and therefore, the introduced errors due to other factors are negligible.
We find an electron density of $n_{\rm e}^{[\rm OII]}$ = 2280$\pm 1900$ cm$^{-3}$, while using $\tt IRAF$ $\tt temden$ task $n_{\rm e}^{[\rm OII]} \sim$  1340 cm$^{-3}$.
The difference of n$_{\rm e}$ as derived from Equation (\ref{eq:3}) and $\tt temden$ is expected, as shown in Figure~1 of \cite{Sanders2016}.
Moreover, we derived very similar electron density, $n_{\rm e}^{[\rm OII]}$=2754 cm$^{-3}$, when using the theoretical relation between [OII]$\lambda$$\lambda$3729/3726 line ratio and electron density from \cite{Kewley2019}, while the ISM pressure derived from the same theoretical relation results to be log(P/k)~$\sim 7.89$.

\begin{figure} 
        \centering
        \includegraphics[width=\linewidth]{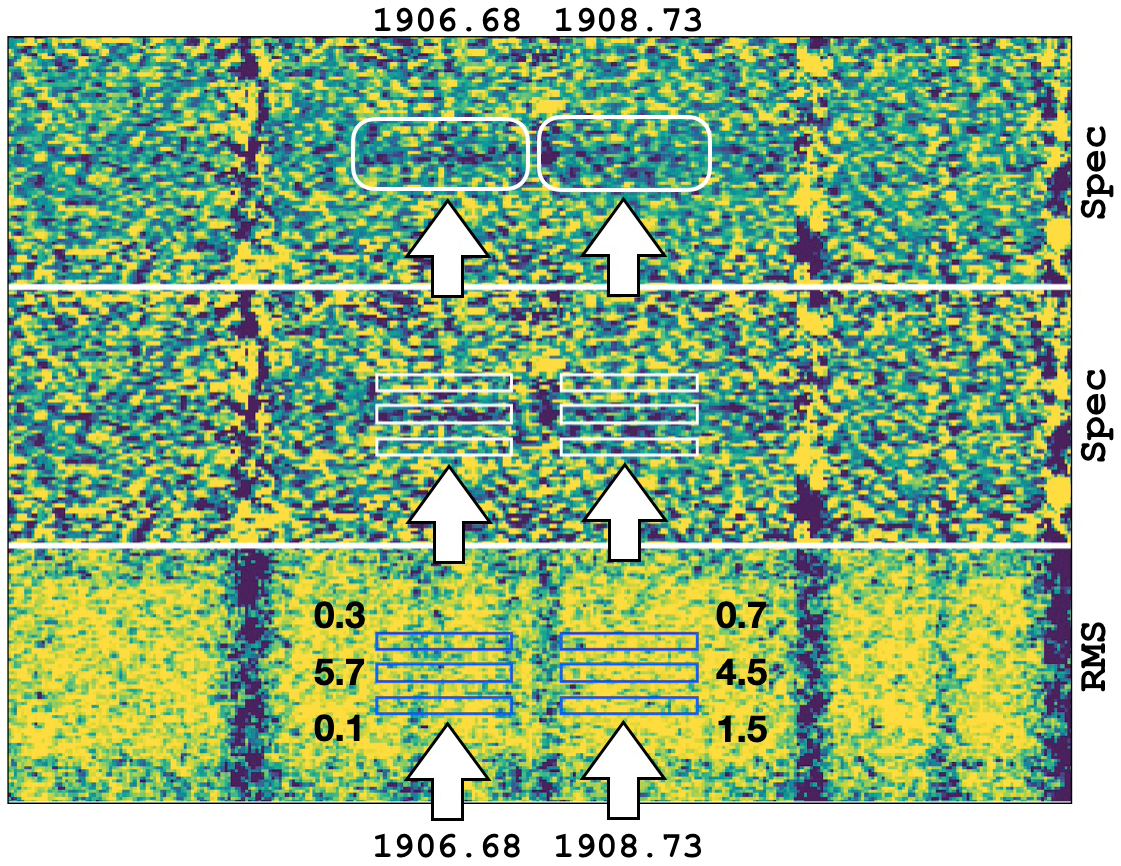}
        \caption{The zoomed X-Shooter spectrum centerd on the CIII] doublet. Top: the two-dimensional spectrum at the original spectral resolution (R=8900, 0.2\AA/pix) with indicated the  positions of the components. Middle: the same spectrum as above with outlined the regions where the signal to noise ratio (SNR) were calculated in three spatial positions: below, above and at the location of Ion3. The size of the boxes is $0.8'' \times 250$~km~s$^{-1}$. Bottom: the two-dimensional RMS spectrum is shown with reported the same boxes shown in the middle panel with listed the corresponding SNR values. Detections emerge on top of Ion3 spatial position. 
        }
        \label{CIII}
\end{figure}

Another density and pressure diagnostic we can use to probe the ISM physical conditions of Ion3 is the [CIII]$\lambda$1909/CIII]$\lambda$1906 line ratio \citep{Kewley2019}.
This ratio is useful for diagnostics in environments characterized by high-pressure log(P/k)>7.5 and high densities, in he range $n_{\rm e}>3\times10^3 - 5.5\times10^5$~cm$^{-3}$.
Therefore it is recognized as a more reliable tracer, for cases where more extreme conditions are present, than the [OII] diagnostic.
The [CIII]$\lambda$1909 and CIII]$\lambda$1906 feature is clearly detected in the binned spectrum shown in Figure~\ref{VIS_spec}. Figure~\ref{CIII}, instead, shows the two-dimensional non-binned zoomed spectrum of the same doublet. The two components CIII]$\lambda$1906 and [CIII]$\lambda$1909  are detected with signal to noise ratios 5.7 and 4.5, respectively, eventually providing a ratio [CIII]$\lambda$1909/CIII]$\lambda$1906$~=~0.7 \pm 0.3$. Ratios lower than 1 suggest extreme ISM conditions, with $n_{\rm e}^{\rm CIII}>10^{4}$cm$^{-3}$ and pressure log(P/k)>7.9.

It is worth noting that such extreme pressure values are typical of the most highly pressurized, turbulent  medium in molecular clouds of local star-forming galaxies \citep{Molina2020, Chevance2020} and high-redshift clumps in high-resolution simulations \citep{Calura2022}.
In virialized systems like star clusters, where equipartition between gravity and kinetic energy holds and where the 
gravitational pressure can be computed as $P_{\rm grav}$= G$ ~ \times ~\Sigma^{2}$ these P/k values correspond to 
gas surface density $\Sigma\sim10^{4} \rm M_\odot$/pc$^{2}$.
Note also that these values are by several orders of magnitude higher than the warm ISM in the Milky Way disc, characterized by typical thermal P/k $\sim 10^{3}\rm K$ $\rm cm^{-3}$.
Overall results from [OII] and indications from [CIII]/CIII] line ratio diagnostics suggest that Ion3 has an extreme and dense ISM environment with high pressure.
This suggests the formation and presence of very massive stars, a conclusion which is also supported by the large ionising photon production efficiency, $\rm log(\xi_{ion}/[Hz~erg^{-1}])=25.5$, as measured in \citet{Ion3_2018} (see also, \citealt{schaerer2025_VMS}).
Deeper spectroscopic observations of CIII features are required to derive precise ratio among the components, however the current analysis suggest a ratio favoring a brighter blue component.

\subsection{Unraveling the ISM ionization state of Ion3?}
\label{sec:5.2}

The established empirical correlation between [NeIII]$\lambda$3870/[OII]$\lambda$$\lambda$3727,29 and [OIII]$\lambda$5007/[OII]$\lambda$$\lambda$3727,29 (O32) line ratios enables us to estimate the O32 index of Ion3, despite the lack of [OIII]$\lambda$5007 wavelength coverage in the data. 
The [NeIII]$\lambda$3968 emission line is detected in the X-Shooter spectrum, while the other component, [NeIII]$\lambda$3870, falls in a strong 
atmospheric absorption band\footnote[1]{https://www.astronomy.ohio-state.edu/pogge.1/Ast161/Unit5/atmos.html}, and is absent in the spectrum.
However, from the theoretical line ratio [NeIII]$\lambda$3870/[NeIII]$\lambda$3968$\sim$2.4 , we estimate the flux of the [NeIII]$\lambda$3870 which results F$_{\rm [NeIII]}\sim 3.74 \times 10^{-17} \rm ~erg/s/cm^2$, 
implying a significant large [NeIII]$\lambda$3870/[OII]$\lambda$$\lambda$3727,29 ratio of 6.9 \citep[][]{Witstok2021}.
Adopting the empirical relationship \citep[Equation 1,][]{Witstok2021}, the inferred line ratio would formally corresponds O32 > 100. 
Conservatively we use the measured NeIII]$\lambda$3968 in place of [NeIII]$\lambda$3870, knowing that in this case the resulting O32 is a lower limit $> 50$.
Still, this value is significantly higher than the limiting ratio of O32$\geq$5, characteristic of most other known LyC leakers at both low and high redshifts \citep{Witstok2021}.
Furthermore, a high O32 line ratio is correlated with a high ionization parameter (log$U$), defined as the ratio of the number density of ionizing photons to the number density of particles. We derived log$U$ using two methods,  following the prescription from \cite{Diaz2000} and using the [NeIII]$\lambda$3870/[OII]$\lambda$$\lambda$3727,29 ratio \citep[Equation 3,][]{Witstok2021}. In both cases, the resulting ionization parameter was log$U>-$1.5, significantly higher than that of other known LyC leakers.  Furthermore, comparing the log$U$ of Ion3 with that of star-forming galaxies at high redshift, located in both lensed and non-lensed fields \citep{Reddy2023, Williams2023, Tang2023}, the value we inferred is remarkably high, comparable to the most extreme cases known in the local and high-z Universe \citep[e.g.,][]{Jaskot2019_large_O32, Topping2024}.
The ongoing intense starburst hosting young massive stars is  consistent with extreme ISM conditions, whose estimated electron density is $n_{\rm e}^{\rm [OII]}$=2280$\pm1900$cm$^{-3}$ and $n_{\rm e}^{\rm CIII]}>10^{4}$cm$^{-3}$. 
However, in environments with high electron density, the strength of low ionization emission features can be suppressed due to increased collisional de-excitation, which can further affect the measured line ratios and inferred ionization parameter.
Either way, both conditions require a hard ionization field generated from young massive stars, responsible for the production of LyC radiation here observed with Hubble data. 

Finally, recent JWST observations of high redshift ($z>6$) sources suggest that these extreme conditions may not be rare. For example, an extreme O32 = 184 is reported at $z>6$ 
by \citet{Topping2024}. This source is recognized as a compact clump with the size of $\sim$20 pc having high SFR surface density, sub-solar metallicity and ISM under extreme ionization conditions with electron density and ionization parameter estimated to be $n_{\rm e}^{\rm CIII]}$=1.1$\times$10${^5}$ and log$U\sim-$1, respectively.
Similar conditions are reported in another spectroscopically confirmed high-redshift galaxy, at $z=12.34$ \citep[GHZ2,][]{Castellano2024_z12, Calabro2024}. This galaxy is characterized by very high stellar and star-formation rate density, $10^{4} - 10^{5}$ $\rm M_\odot\,pc^{-2}$ and $\Sigma_{\rm SFR} \sim 10^{2} - 10^{3}~\rm M_\odot$~yr$^{-1}$~kpc$^{-2}$, respectively.
Furthermore, the ISM of GHZ2 is characterized by low metallicity,  $<0.1Z_\odot$, high ionizing parameter, log$U>-2$, and extreme ionizing conditions, with O32$\sim$25. Ion3 shares similar physical properties of the ISM, observed during a young bursty event.

\subsection{Star formation rate}
\label{sec:5.3}
Another key physical property that can be extracted from the forbidden [OII]$\lambda$$\lambda$3727,3729 line (in this case the dust-uncorrected line flux) is the star-formation rate (SFR), estimated as:

\begin{equation}
\label{eq:4}
\rm SFR\,(\rm M_\odot\,yr^{-1})=(1.4\pm0.4)\times10^{-41}\,L[OII](erg\,s^{-1}),
\end{equation}

where L[OII] is luminosity of the line.
This prescription to derive the SFR from the [OII] forbidden line is adopted from \cite{Kennicutt1998} and has been calibrated using the H$\alpha$ line from two spectro-photometrical samples of galaxies \citep{Gallagher1989, Kennicutt1992}.
\cite{Gilbank2010} introduced an empirically-corrected SFR that takes into account metallicity and dust obscuration on [OII] as a function of galaxy mass:

\begin{equation}
\label{eq:5}
\rm SFR([OII])_{corr} = \frac{\rm SFR_{[\rm OII]}}{a\, \rm tanh[(x-b)/c]+d},
\end{equation}

where SFR$_{\rm [OII]}$=L([OII])/2.53$\times 10^{40}\rm erg s^{-1}$, x=log(M$_{\ast}$/\rm M$_\odot$), a=--1.424, b=9.827, c=0.572 and d=1.700.
After adopting the above-mentioned empirical correction, we find  SFR([OII])$_{\rm corr}$ = 12$\pm5$ $\rm M_\odot\,yr^{-1}$. Lastly such way deriven SFR([OII]) was additionaly corrected by total magnification of 1.3, resulting in final SFR([OII])$_{\rm corr,}^{\rm tot}$ = 9$\pm4$ $\rm M_\odot\,yr^{-1}$.

However, the [OII] SFR indicator is prone to substantial systematic uncertainties caused by different galaxy properties \citep[ionization parameter, metallicity, extinction, etc][]{Kennicutt1998, Teplitz2003, Kewley_2004_[OII], Moustakas2006}, and differences in the SFR inferred from [OII] and H$\alpha$ have been found to be as high as 0.5--1 dex \citep{Kennicutt1998, Kewley_2004_[OII]}, namely SFR([OII])$_{\rm corr}$ can be subjected the large systematics.

On the other hand, the rest frame UV non-ionizing continuum is produced from stars with a broader mass range, extending down to a few M$_\odot$.
This implies that rest-frame UV flux traces SFR events over larger time periods of about 100 Myr.
The inferred SFR(UV) from the F814W band probing the non-ionizing ultraviolet luminosity is SFR(UV)=10$\pm1$~$\rm M_\odot\,yr^{-1}$ \citep{kenni12}, where no dust attenuation is considered.
To correct SFR(UV) from dust extinction we applied the correction using A1600 -- $\beta$ relation: A$_{1600}$= 5.32$^{+0.41} _{-0.37} +1.99 \times \beta$ from \cite{Castellano2014}, adopting $\beta$=--2.50 (\citealt{Ion3_2018}). The dust-corrected SFR is SFR(UV)$_{\rm corr}\sim$14~$\rm M_\odot\,yr^{-1}$ and after applying magnification correction it decreases to  $\sim11$~$\rm M_\odot\,yr^{-1}$.

A more reliable SFR tracer accessible for Ion3 is the H$\alpha$ line. As discussed in \citet{Ion3_2018} Spitzer/IRAC photometry shows significant photometric excess in the 3.6µm band, which, at its redshift, includes the H$\alpha$ line. We derive the SFR(H$\alpha$) attributing this excess to H$\alpha$ line. 
From the magnitude of the galaxy in the IRAC 3.6 µm band, mag=23.20$\pm0.10$ and subtracting the continuum, guessed from the nearest band, IRAC 4.5 µm, to be mag=23.70$\pm0.10$, we derive an H$\alpha$ line flux $F_{\rm H\alpha}$=1.2$\pm0.4$$\times10^{-16}\rm erg/s/cm^2$.
The resulting SFR(H$\alpha$) is $100\pm30$ $\rm M_\odot\,yr^{-1}$ \citep{Kennicutt2012} and after correcting for magnification it is SFR(H$\alpha$)$^{\rm tot}$ = $77\pm23$ $\rm M_\odot\,yr^{-1}$. The value presented here is slightly lower than that reported by \citet{Ion3_2018} (140 M$_\odot~\text{yr}^{-1}$), primarily because we correct for continuum emission and assume that the continuum level of 23.70 also applies to the IRAC 3.6 µm channel.

H$\alpha$ is a nebular emission line produced by the recombination of gas ionized from short-living young and massive stars. 
Therefore, H$\alpha$ is an excellent tracer of SFR events that span over short periods ($<10$~Myr). 
The relatively high SFR(H$\alpha$) indicates  Ion3 is undergoing an intense burst of star-formation, in line with the detected LyC emission, steep ultraviolet slope and  signatures of massive stars in the ultraviolet spectrum (e.g., P-Cygni profile of NV). At the give stellar mass of $\sim$1.5 $\times 10^{9}$ M$_\odot $, the  SFR(H$\alpha$) places the LyC emitter Ion3 in the starburst phase, with specific star formation rate of $\simeq 70$~Gyr$^{-1}$, fully compatible with the strong H$\alpha$ emitters observed during the epoch or reionization \citep[e.g.,][]{Rinaldi2024_bursty_LyC}. It worth reporting also the star formation rate surface density, $\Sigma_{\rm SFR}$, calculated as the ratio between the SFR(H$\alpha$) and the area outlined by the two-sigma contour defined in the F814W band (which probes the rest-frame non-ionizing ultraviolet radiation $\sim 1600$\AA). Assuming the bulk of the SFR activity is confined within that area, corresponding to 5.5~kpc$^{2}$ (see Figure~\ref{gal}), $\Sigma_{\rm SFR}$ is $\simeq 20$~M$_\odot$~yr$^{-1}$~kpc$^{-2}$. This high value is consistent with the  high electron density inferred above \citep[][see also \citealt{Topping2025_ne}]{Reddy2023_SigmaSFR_Ne, Reddy2023}.

\subsection{The CIII]$\lambda$$\lambda$1907,1909 metallicity}
\label{sec:5.4}

The CIII]$\lambda$$\lambda$1907,1909 (CIII]$\lambda$1909) is a very useful spectral feature to identify very distant star-forming galaxies, e.g. at $z>6$, especially in the pre-JWST era.
\citep[e.g.,][]{Shapley2003, Stark2014, Rigby2015, Ravindranath2020}.
Moreover, CIII]$\lambda$1909 proves to be a decent metallicity diagnostic for low mass star-forming galaxies down to 12+log(O/H)$\sim7.5$ \citep[e.g.,][]{Rigby2015, Nakajima2018, Ravindranath2020, Mingozzi2022}, and it is also considered as a potential tracer of LyC leaking galaxies, as shown for "Grean Pea" galaxies \citep[][]{Cardamone2009} with detected LyC emission \citep{Schaerer2022, Izotov2023}.

To evaluate the metallicity of Ion3, we are using the best fitting results of estimated gas-phase metallicities and measured (CIII]$\lambda$1909 from the sample of local high-$z$ analogs.
Which is proven to be valid at 12+log(O/H)>7.5 with a scatter of 0.18dex \citep{Mingozzi2022}:

\begin{equation}
\label{eq:6}
12+\rm log(O/H) = (-0.5\pm0.13) \times log(EW(CIII]))+(8.43\pm0.11),
\end{equation}

\noindent where EW(CIII]) is the rest-frame equivalent width of the doublet. The inferred EW(CIII])=6.5Å reported in Section \ref{sec:3.2.1} implies 
a metallicity 12+log(O/H)=8.02$\pm0.20$ or $\sim$0.2Z$_{\odot}$. 
This sub-solar metallicity agrees with the measured values of other confirmed LyC leakers at high redshift, such as as Ion2 \citep[12+log(O/H)=8.07$\pm0.44$,][]{vanz_ion2, deBaros2016_ion2} and Sunburst \citep[12+log(O/H)$\sim$8.4,][]{rivera17, vanz_sunburst, chisholm19, Mainali2022, Mestric2023}.

\begin{table}
\caption{Resulting photometry, spectroscopy and physical properties of Ion3.}
\label{tab:1}
\begin{tabular}{cccc}
\multicolumn{4}{c}{Photometry (sources A+B)}    \\ \hline \hline
band    & mag     & error   & SNR   \\ \hline
F390W   & 28.80   & 0.30       & 3.5   \\ 
F814W   & 24.20   & 0.07       & 15    \\ \hline
\multicolumn{4}{c}{Spectroscopy}       \\ \hline \hline
line                 & \begin{tabular}[c]{@{}c@{}}$\lambda_{rest}$\\ {[}Å{]}\end{tabular} & \begin{tabular}[c]{@{}c@{}}flux×10$^{-17}$\\ {[}erg/s/cm$^{2}${]}\end{tabular} & \begin{tabular}[c]{@{}c@{}}error×10$^{-17}$\\ {[}erg/s/cm$^{2}${]}\end{tabular} \\ \hline
HeII         & 1640.42       & 0.34       & 0.25   \\
{[}OII{]}    & 3727.09       & 0.34       & 0.17   \\
{[}OII{]}    & 3729.88       & 0.21       & 0.14   \\
{[}NeIII{]}  & 3968.53       & 1.56       & 0.54   \\ \hline
\multicolumn{4}{c}{Physical properties (sources A+B)}         \\ \hline \hline
                         & value                & error {[}$\pm${]}     & units   \\ \hline
$n_{\rm e} ^{\rm [OII]}$ & 2280    & 1900         & cm$^{-3}$     \\
$n_{\rm e} ^{\rm CIII]}$ & \textgreater{}10$^{4}$    & -     & cm$^{-3}$   \\
O32                      & \textgreater{}100         & -     & -           \\
SFR(H$\alpha$)$^{\rm tot}$           & 77       & 23        & $\rm M_\odot\,yr^{-1}$  \\
SFR([OII])$_{\rm corr}^{\rm tot}$       &  9    & $\sim$4   & $\rm M_\odot\,yr^{-1}$              \\
SFR(UV)$_{\rm corr}^{\rm tot}$              & $\sim$11    & -     & $\rm M_\odot\,yr^{-1}$              \\
$\Sigma_{\rm SFR}$                &     $\simeq 20$    &   -     &   $\rm M_\odot$~yr$^{-1}$~kpc$^{-2}$ \\
12+log(O/H)          & 8.02  & 0.20  & -                                   \\
$f_{\rm esc,rel}$    & 0.06 - 1      & -          & -                      \\ \hline             \multicolumn{4}{c}{Morphology}         \\ \hline \hline
                         & Size                & error {[}$\pm${]}     & units   \\ \hline
Ion3$\rm _A$        &        180       &           90     &              pc \\
Ion3$\rm _B$        &       <100      &             -    &                pc \\ \hline

\end{tabular}
\end{table}

\subsection{Complexity of the  Ly$\alpha$ line}
\label{sec:5.5}

Interpreting the properties of the Ly$\alpha$ line is challenging due to its resonant nature. 
The way how Ly$\alpha$ photons propagates through the ISM and CGM is affected by the intrinsic properties of the medium (e.g., geometry, ionization and kinematics of the gas, dust attenuation, etc.), and leaves an imprint in the shape 
of the line.
Therefore, Ly$\alpha$ is a carrier of valuable information about various intrinsic properties of the neutral hydrogen HI of the hosting galaxy.
As a consequence of its nature, the Ly$\alpha$ profile is recognized as one of the most robust indirect tracers of escaping LyC photons \citep{Verhamme2015}, which is later proven by confirmed LyC emitters at various redshifts having complex Ly$\alpha$ profiles \citep[][]{Verhamme2017, izotov21loz_analogs, vanz_sunburst, Naidu2022} (but see also the Ion1 LyC emitter at z=3.8 without Ly$\alpha$ emission, \citealt{Ion1_Zhiyuan_2020}).

\begin{figure} 
        \centering
        \includegraphics[width=\linewidth]{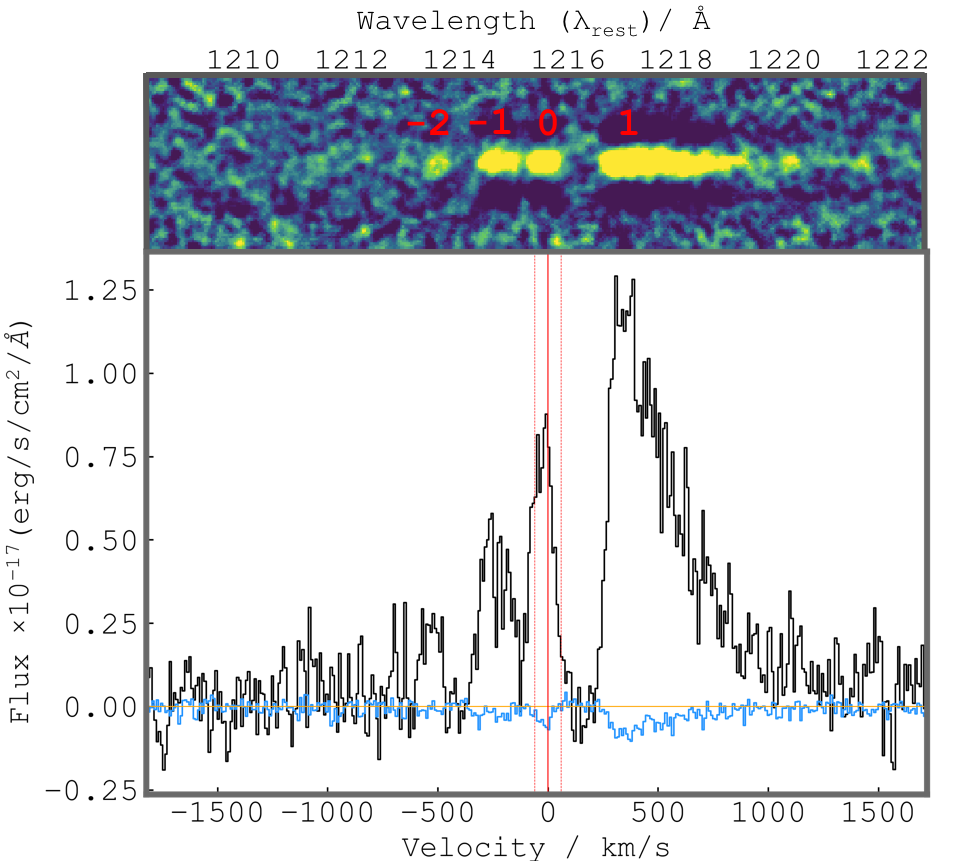}
        \caption{Upper and bottom panels show the two-dimensional (2D) and the one-dimensional (1D) spectra of the Ly$\alpha$ extracted from X-Shooter. The blue line shows the spectrum of the sky, while the red vertical line (on the 1D spectrum) marks the systemic velocity of the source with the one sigma error (dotted vertical lines), located at +35 km~$\rm s^{-1}$ form the central Ly$\alpha$ peak dubbed 0. The other Ly$\alpha$ peaks are marked as -2, -1 (blueward the systemic velocity) and 1 (redward from systemic velocity)}.
        \label{lya}
\end{figure}

Figure \ref{lya} shows the complex Ly$\alpha$ multi-peaked shape, in which a clear narrow emission close to the systemic redshift (red line) is evident  (Ly$\alpha$~$-$~systemic $= dv \sim$ $-35$ km~$\rm s^{-1}$).
In particular, including the uncertainty of the systemic redshift $3.999\pm0.001$, $dv$ ranges from $-60$ to $+10$ km~$\rm s^{-1}$, making the peak fully compatible with being at the systemic (dotted red lines in the same figure).
The proximity of Ly$\alpha$ line location with its resonance frequency strongly indicates very low column density of neutral HI gas \citep[e.g.,][]{rivera17, behrens14, Naidu2022}, in line with the LyC detection reported in this work and suggesting the presence of ionized channels through which Ly$\alpha$ and LyC photons can escape into the IGM \citep{Verhamme2017}. It is worth noting that besides the presence of aligned ionized channels, there are ongoing radiative transfer effects that produce additional peaks observed far from the resonance frequency. A characteristic also observed in other confirmed leakers, like the {\rm Sunburst} arc at z=2.37 \citep[e.g.,][]{rivera17}, Ion2 at z=3.2 \citep[e.g.,][]{vanz_ion2} and  among the local analogs, remarkably similar to the case of triple-peaked Ly$\alpha$ emitter J1243+4646 \citep[][]{Izotov2018_LyC_0.46}.

\section{Summary and Conclusions}
\label{sec:6}

In this work, we present detailed HST multi-band (F390W, F814W, and F140W) photometric and high-resolution VLT X-Shooter spectroscopic analysis of the most distant LyC leaker at $z_{spec}$=3.999, summarized in Table \ref{tab:1}.
The HST F390W band covers a clean non-contaminated portion of the LyC flux blueward $\sim880$Å while F814W covers the UV non-ionizing part $\sim$1500Å.
Additional ancillary X-Shooter spectroscopy allowed us to investigate properties of spectroscopic features focusing on $\sim$ CIV$\lambda$1550, HII$\lambda$1640, CIII]$\lambda$$\lambda$1907,1909, NeIII]$\lambda$3968 and [OII]$\lambda$$\lambda$3726,3729.
Our main results are summarized as follows:

\begin{enumerate}

\item We measured a non-contaminated LyC radiation using HST F390W band with SNR$\sim$3.5, which results in $f_{\rm esc,rel}$=0.06--1, depending on the $\tau_{IGM}^{LyC}$. 
\item Additional spectral features have been identified in the X-Shooter spectrum and confirmed the systemic redshift z$_{\rm spec}$=3.999$\pm0.001$, based on [OII]$\lambda$$\lambda$3726,3729 and NeIII]$\lambda$3968 nebular features, revealing that within the uncertainties the Ly$\alpha$ central peak lies at the systemic velocity z$_{\rm sys}$.
Furthermore, Ion3 shows signatures of an ongoing burst (P-Cygni profile of NV, rest frame EW(H$\alpha$)$\sim$1000Å and high [NeIII]$\lambda$3870/[OII]$\lambda$$\lambda$3727,29 implying a large O32>50.
\item We find high electron density,  $n_{\rm e}^{\rm [OII]}$=2280$\pm1900$ cm$^{-3}$, indications for $n_{\rm e}^{\rm CIII]}>10^{4}$ cm$^{-3}$ and sub-solar metallicity 12+log(O/H)=8.02$\pm0.20$, suggesting extreme ISM conditions connected with the high star-formation rate surface density. 
\item The star formation rate (SFR), has been evaluated, and corrected for magnification, using three different diagnostics ([OII], H$\alpha$, UV), yielded the following results: SFR([OII]) = 9$\pm$ 4 $\rm M_\odot yr^{-1}$, SFR(H$\alpha$)$^{\rm tot}$ = 77 $\pm$ 23 $\rm M_\odot yr^{-1}$, and SFR(UV)$_{\rm corr} \sim$ 11 $\rm M_\odot yr^{-1}$. The star formation rate surface density of Ion3 is relatively high $\Sigma_{\rm SFR} \simeq 20$~M$_\odot$~yr$^{-1}$~kpc$^{-2}$ and in line with the high electron density inferred in Sect.~\ref{sec:5.1}.
\item High-resolution HST imaging reveals the clumpy structure of Ion3.
Using the F814W image, we measure the effective radius of Ion3$_{\rm A}$ $\sim$ 180 pc, while an upper limit is derived for the second component, Ion3$_{\rm B}$<100 pc, while the currently estimated area in the rest-frame 1600\AA~is 5.5~kpc$^{2}$.

\end{enumerate}

\begin{acknowledgements}
The authors thank the anonymous referee for helpful comments that improved the manuscript. We thank Rui Marques-Chaves on helpful discussion related to $\tau_{IGM}^{LyC}$.
The analysis presented in this paper is based on observations with the NASA/ESA Hubble Space Telescope obtained at the Space Telescope Science Institute, which is operated by the Association of Universities for Research in Astronomy, Incorporated, under NASA contract NAS5-26555. Support for Program Number HST-GO-17133 was provided through a grant from the STScI under NASA contract NAS5-26555.
UM, CG acknowledge financial support through grant PRIN-MUR 2020SKSTHZ
M.M. acknowledges the financial support through grant PRIN-MIUR 2020SKSTHZ.
MC acknowledges support from INAF Mini-grant ``Reionization and Fundamental Cosmology with High-Redshift Galaxies."
AZ: The research activities described in this paper have been co-funded by the European Union – NextGenerationEU within PRIN 2022 project n.20229YBSAN - Globular clusters in cosmological simulations and in lensed fields: from their birth to the present epoch.
AM acknowledges financial support through grant NextGenerationEU" RFF M4C2 1.1 PRIN 2022 project 2022ZSL4BL INSIGHT.
KIC acknowledges funding from the Dutch Research Council (NWO) through the award of the Vici Grant VI.C.212.036.
This research made use of Photutils, an Astropy package for
detection and photometry of astronomical sources \citep{Photutils2024}.
This work uses the following software packages:
\href{https://github.com/astropy/astropy}{\texttt{Astropy}}
\citep{astropy1, astropy2},
\href{https://github.com/matplotlib/matplotlib}{\texttt{matplotlib}}
\citep{matplotlib},
\href{https://github.com/numpy/numpy}{\texttt{NumPy}}
\citep{numpy1, numpy2},
\href{https://www.python.org/}{\texttt{Python}}
\citep{python},
\href{https://github.com/scipy/scipy}{\texttt{Scipy}}
\citep{scipy}.
\end{acknowledgements}

\bibliographystyle{aa} 
\bibliography{refs} 

\begin{appendix} 

\end{appendix}

\end{document}